\documentclass[twocolumn,nofootinbib,superscriptaddress,preprintnumbers,floatfix]{revtex4-1}
\usepackage{amsmath}
\usepackage{amsfonts}
\usepackage{amssymb}
\usepackage{graphicx}
\usepackage{epsfig}
\usepackage{subfigure}
\usepackage{color}
\usepackage{float}

\newcommand{\be}{\begin{equation}}
\newcommand{\ee}{\end{equation}}
\newcommand{\ba}{\begin{eqnarray}}
\newcommand{\ea}{\end{eqnarray}}

\def\slash#1{#1 \hskip-0.45em /}
\def\Li{\textrm{Li}}
\def\ln{\textrm{ln}}

\def\nn{\nonumber}

\def\se{{\varepsilon}}

\begin{document}

%%%%%%%%%%%%%%%%%%%%%%%%%%%%%%%%%%%%%%%%%%%%%%%%%%%%%%%%%%%%%%%%%%%%%%%%%%%%%%%%
% Title page
%%%%%%%%%%%%%%%%%%%%%%%%%%%%%%%%%%%%%%%%%%%%%%%%%%%%%%%%%%%%%%%%%%%%%%%%%%%%%%%%

\preprint{\vbox{\hbox{UWTHPH 2012-20}
%\hbox{\today}
}}

%\vfill
\title{\Large Secondary Heavy Quark Production in Jets through Mass Modes}

\author{ Simon Gritschacher}
\affiliation{Mathematisches Institut, Ludwig-Maximilians-Universit\"at M\"unchen
Theresienstra\ss{}e 39, 80333 M\"unchen, Germany}

\author{ Andre H. Hoang, Ilaria Jemos and Piotr Pietrulewicz}
\affiliation{Fakult\"at f\"ur Physik, Universit\"at Wien,
Boltzmanngasse 5, 1090 Vienna, Austria}

\begin{abstract}
We present an effective field theory method to determine secondary massive quark
effects in jet production taking the thrust distribution for  $e^+ e^-$ collisions in the dijet limit as a concrete
example. The method is based on the field theoretic treatment of collinear and
soft mass modes which have to be separated coherently from the collinear and
ultrasoft modes related to massless quarks and gluons. For thrust the structure
of the conceptual setup is closely related to the production of massive gauge
bosons and involves four different effective field theories to describe all
possible kinematic situations. The effective field theories merge into each
other continuously and thus allow for a continuous description from infinitely heavy to arbitrarily
small masses keeping the exact mass dependence of the most singular terms
treated through factorization. The mass mode field theory
method we present here is in the spirit of the variable fermion number scheme
originally proposed by  Aivazis, Collins, Olness and Tung and can also be 
applied in hadron collisions.   
\end{abstract}
%\vfill
%{\bf PACS:}
%\vfill

\maketitle

\section{Introduction}\label{sect:intro}

By now jet physics has reached a high level of precision allowing for
an accurate description of the strong interaction. Fundamental parameters of
QCD such as the strong coupling constant as well as
nonperturbative properties of hadrons like parton distribution functions can be
determined with a continuously improved accuracy from high precision data
samples. On the theoretical side this has been possible thanks to high
order loop calculations and the summation of large logarithmic terms.
Computations of jet cross sections for massless quarks belong to the well known
and unambiguously defined exercises in perturbative QCD based on a number of
rigorous  factorization proofs. On the other hand, as far as massive quarks are
concerned it is fair to say that their treatment is not coherent throughout the
literature. Different schemes for massive quarks exist which differ in
the resummation of logarithms and in the inclusion of formally
subleading contributions. An approach capable of describing quark mass effects
starting in principle from very small masses when the quarks are inside hadrons
and stretching up to ultra-heavy masses in the decoupling limit was provided by
Aivazis, Collins, Olness and Tung
(ACOT)~\cite{Aivazis:1993pi,Aivazis:1993kh}. Their variable fermion 
number scheme is based on the separation of close-to-mass-shell and off-shell
modes. At LO in the inverse hard scattering scale expansion, it allows to factorize infrared-safe hard coefficient corrections
 from
low-energy parton distribution terms involving logarithmic mass effects. In this
respect the concept behind the ACOT scheme is along the lines of effective field
theory methods such as the soft-collinear effective theory (SCET)
framework~\cite{Bauer:2000ew, Bauer:2000yr} and can be readily incorporated into
it. As we show in this paper, the resulting effective theory framework can be based
on the inclusion of collinear and soft {\it "mass~modes"}\footnote{This differs
  from the   terminology used in Refs. \cite{Fleming:2007qr} and
  \cite{Fleming:2007xt} where only the soft massive modes were referred to as
  the mass modes.}   
together with the existing collinear and ultrasoft massless partonic modes. While the
collinear and ultrasoft massless partonic modes typically have different invariant
masses depending on the observable under consideration, the collinear and soft
mass modes also contain fluctuations related to their intrinsic mass scale. This can lead to complicated
patterns of scale hierarchies that might even vary substantially within a single
distribution.

\begin{figure}[t]
 \centering
 \includegraphics[width=7cm]{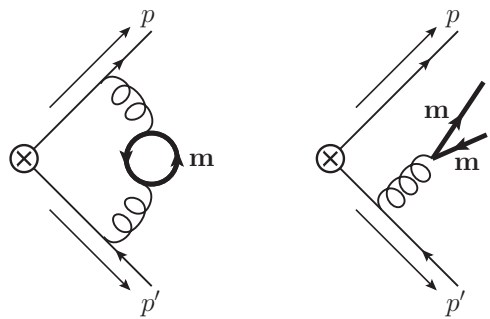}
 \caption{Diagrams at $\mathcal{O}(\alpha_s^2)$ for virtual and real secondary
   radiation of massive quark pairs in primary massless quark
   production.\label{fig:QCDdiag2}} 
\end{figure}
In this and a subsequent paper we apply the mass mode SCET concept to describe
the {\it secondary} production of massive quarks in the $e^+e^-$ thrust
distribution. While we aim for the description of thrust, the described method
is general and can be applied with possible adaptions to other processes as
well. 
Recently, a factorization approach was derived within SCET that can be
applied for thrust induced by {\it primary} heavy quarks, i.e.\ by heavy quarks
that are produced directly from the hard jet
current~\cite{Fleming:2007qr,Fleming:2007xt}. For the description of  {\it
  primary} heavy quarks the mass modes are involved as well, but they play a
more simplistic role since massless collinear modes are directly tied to the
massive collinear modes and soft mass modes turn out to only lead to virtual
effects. So the resulting situation is almost identical to the one in massless
SCET. For {\it secondary} heavy quark production, i.e.\ for heavy quarks
produced from gluon radiation off primary massless quarks,
Fig.~\ref{fig:QCDdiag2}, the collinear and ultrasoft massless as well as massive
modes can all constitute independent degrees of freedom. Each of them can
  contribute to
dynamical thresholds effects as well as to virtual contributions. Here the dynamical mass effects enter in
different ways depending on the relation of the mass scale to the kinematic
scales relevant for massless quark production. Depending on the mass value the
scale hierarchies can vary substantially requiring different types of effective
field theory setups that have to merge into each other in a connected way to allow
for a continuous description of the distribution. The secondary quark mass
effects in thrust thus represent a non-trivial  prototypical 
showcase for the mass mode concept.

Concerning the numerical size, the effects from secondary heavy quarks in thrust
are ${\cal O}(\alpha_s^2)$ in 
fixed-order perturbation theory and certainly small if the hard scale $Q$ is much
bigger than the heavy quark mass. Nevertheless the numerical effects involve
logarithms related to renormalization evolution with variable flavor
number already at the leading-logarithmic level and are important if the
hard scale is only a few times larger than the mass. They are certainly essential
for a precision analysis of $e^+e^-$ event shapes for $Q<35$~GeV concerning the
description of bottom
mass effects. So studying the secondary heavy quark effects in
thrust is also of phenomenological interest.

An interesting conceptual issue, which can also lead to considerable
computational simplifications, is that the description of secondary massive
quarks radiation is closely tied to radiation of massive gauge bosons. All
non-trivial conceptual aspects and also some computational issues can be
conveniently addressed considering the setup with massive gauge bosons which we mostly concentrate on
in this first paper. Details on the computation of the effects of secondary
massive quarks and extensive numerical studies are postponed to
Ref.~\cite{Hoang:2013}. 

This paper is organized as follows. In Sec.~\ref{sect:outline} we explain the
outline and the scope of the mass mode method. We also show how secondary
massive quark radiation is tied up to the description of
massive gauge bosons. The required effective field theory framework can be
conveniently formulated concentrating on collinear and soft massive gauge boson
radiation as a placeholder. Essentially all non-trivial conceptual and technical
issues can be discussed and handled with massive gauge bosons. 
In Sec.~\ref{sect:massless} we review the
massless factorization
theorem for the thrust differential cross section. In
Sec.~\ref{sect:setup}, which is the central section of this paper, we
explain our theoretical setup to describe 
modifications in the factorization theorem due to secondary massive
particles described by mass modes, and we show the
results for massive gauge bosons with vector coupling. Here we also address
how our approach achieves a continuous description of the full
mass-dependence of the most singular terms in the dijet limit covering
all possible hierarchies, and we discuss
the role of mass corrections at the transition
points between the different scenarios. Details on the actual calculations are given in
Sec.~\ref{sect:fulltheoryresults} for the full theory results and in
Sec.~\ref{sect:eftresults} in the effective theory framework. As an outlook
we describe briefly the two-loop computations required for secondary massive quarks in
Sec.~\ref{sect:outlook} and show the impact of our results on the thrust
distribution. Finally in Sec.~\ref{sect:conclusions} we summarize and conclude.

\section{The Outline}\label{sect:outline}

We consider the effect of secondary massive quarks in the
$e^{+}e^{-}$ thrust distribution for massless primary quark production,
$d\sigma^{\rm  light}/d\tau$. For thrust we use the definition 
\begin{eqnarray}
\tau=1-T \,=\, 1-\sum_i \frac{ |\mathbf{n} \cdot
\vec{p}_i|}{\sum_j |E_j|}
\,=\, 1-\sum_i \frac{ |\mathbf{n} \cdot
\vec{p}_i|}{Q}\,,
\end{eqnarray}
where $\mathbf{n}$ is the thrust axis and the
sum is performed over all final state particles with momenta $\vec{p}_i$ and
energies $E_i$.\footnote{
We define the thrust variable $\tau$ normalized with the c.m.\ energy $Q$, which is the sum of all
energies and also agrees with the variable 2-jettiness~\cite{Stewart:2010tn}. 
For massless decay products this agrees with the common definition, which is normalized to the sum of momenta $\sum_i |\vec{p}_i|$.}
The most singular terms for small $\tau$ (i.e.\ in the dijet limit) are
governed by a factorization theorem that separates the dynamical fluctuations
at the center-of-mass energy $Q$ (hard scale), the typical invariant mass of
each of the jets $Q \lambda$ (jet scale) and the typical energy of large angle
ultrasoft radiation $Q \lambda^2$ (ultrasoft scale), where the power counting parameter
$\lambda\sim \sqrt{\tau}$ in the tail region of the thrust distribution. For
$m \gg Q$ the effects of the massive quark decouple and the factorization
theorem for the singular partonic cross-section adopts the well known form for
massless quarks~\cite{Fleming:2007qr}. Schematically the form is
\be
\frac{d\sigma}{d\tau} \simeq \mathcal{H}\cdot \mathcal{J} \otimes \mathcal{S} \, .
\label{eq:diffsigmaschem}
\ee
Here  $\mathcal{H}$ denotes the hard contribution, $\mathcal{J}$ the
jet function describing collinear fluctuations, which is convoluted with the soft
function $\mathcal{S}$ arising from soft large angle radiation. To 
avoid large logarithms each term is evaluated at its
characteristic renormalization scale, so that $\mu_H \sim Q$ for the hard factor,
$\mu_J \sim Q\lambda \sim Q\sqrt{\tau}$ for the collinear jet function and
$\mu_S \sim Q\lambda^2 \sim Q\tau$ for the soft function. Large logarithms
between these  scales are summed by the renormalization group factors which are
implied. 

\begin{figure}[t]
 \centering
 \includegraphics[width=1\linewidth]{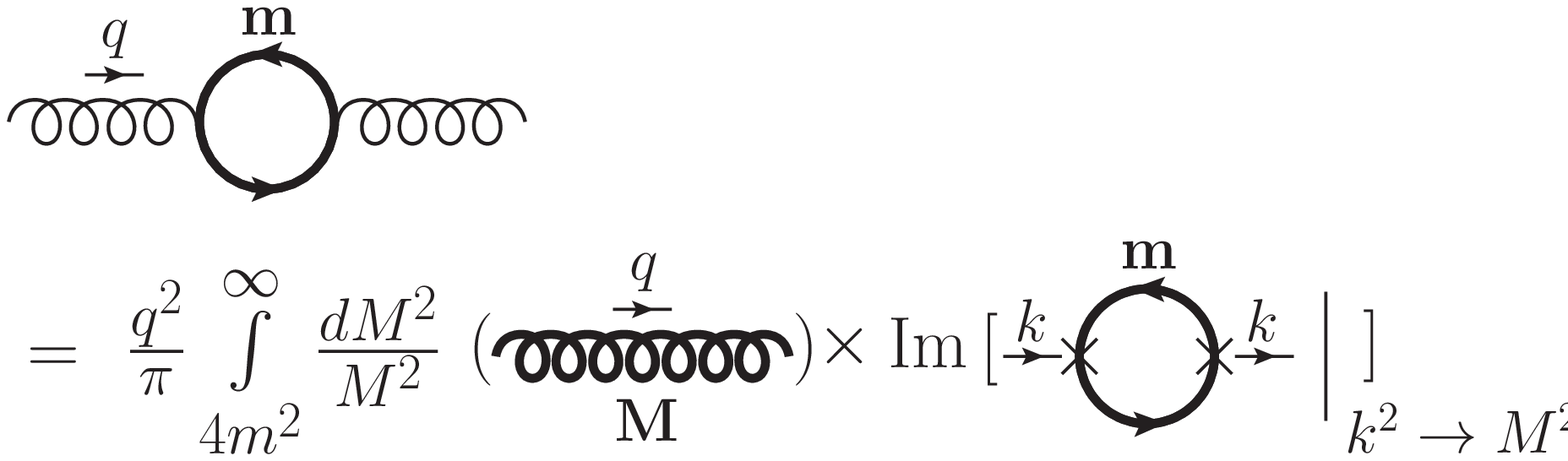}
 \caption{Figure illustrating the dispersion method for the vacuum polarization
   correction to the gluon propagator in the subtracted version with
   $\Pi(q^2=0)=0$ suitable for situations where the massive quark is not
   contributing to the renormalization group evolution. The explicit analytic
   form of the dispersion relations is 
   discussed in Sec.~\ref{sect:outlook}.
   \label{fig:dispersion}} 
\end{figure}
\begin{figure}
 \centering
 \includegraphics[width=7cm]{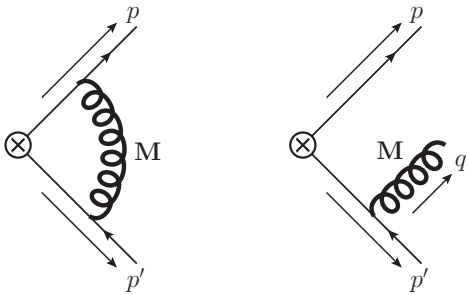}
 \caption{Diagrams at $\mathcal{O}(\alpha_s^2)$ for virtual and real secondary
   radiation of gluons with mass $M$ in primary massless quark
   production. \label{fig:QCDdiag1}} 
\end{figure}

In order to determine the effects of secondary radiation of heavy quarks through
gluon splitting (Fig.~\ref{fig:QCDdiag2}) one has to deal with two issues: (i)
the separation of dynamical modes depending on the relation of the mass $m$ with
respect to the hard, jet and ultrasoft scales and (ii) the calculation of the
secondary massive quark contributions in connection with the appropriate number of
flavors contributing to the renormalization group evolution. For
jet observables and quantities depending only on the invariant mass of the secondary
fermion pair the two issues can be completely disentangled. This is achieved
by using the fact that the quark pair polarization correction to the gluon
propagator can be expressed as a dispersion integral of a massive gluon
propagator with the absorptive part of the quark-vector current correlator (see
Fig.~\ref{fig:dispersion}). Using this dispersion integral
method~\cite{Kniehl:1988id,Hoang:1994it,Hoang:1995ex} 
we can set up the conceptual formalism to describe the secondary massive
quark pair radiation and to separate the massive modes in connection to the
massless collinear and soft modes by first considering real and virtual
radiation of ``gluons'' with mass $M$, as shown in Fig.~\ref{fig:QCDdiag1}. 
All conceptual issues concerning the separation of modes, the determination of
matching conditions and of matrix elements within the context of SCET can be
conveniently discussed with the massive gauge bosons.

One has to consider collinear (for each jet direction)
as well as soft ``gluon'' modes with mass $M$, all of which can
  have mass-shell fluctuations of equal typical virtuality of $\mathcal{O}(M)$ in addition to
  fluctuations at the hard, jet and ultrasoft scale depending on the kinematic situation. 
  In contrast, massless collinear and ultrasoft modes live exclusively
    at the jet and ultrasoft scale, respectively, and their (zero) mass-shell fluctuations do not need to be
    considered for the separation of modes. 
After having carried out all ${\cal O}(\alpha_s)$ calculations with the massive ``gluons'',
the ${\cal O}(\alpha_s^2)$ effects from secondary massive quark radiation are obtained
by the dispersion integration over the gauge boson mass according to
Fig.~\ref{fig:dispersion}. The dispersion integration can be carried out either in the
unsubtracted or subtracted versions depending on whether the massive quark flavor
gives a UV divergence and contributes to the RG-evolution or whether it is
integrated out and does not contribute to the RG-evolution.

In this paper we mostly concentrate on the conceptual aspects related to the
collinear and soft gauge bosons with mass $M$ and on their ${\cal O}(\alpha_s)$
effects in thrust. We aim at a continuous
description of the most singular terms treated in the factorization
  theorem for all mass values from $M \gg Q$, where decoupling takes place,
down to the limit $M\rightarrow 0$, where the formalism continuously interpolates to
the well-known case of a massless gluon. The linearity of the dispersion
integration entails that the continuity is carried over also to the
contributions coming from the secondary massive quarks.  
The continuity of the theoretical description as a function of $Q$, the mass and
the external kinematical variables (which is thrust $\tau$ in our concrete
application) is particularly important, since 
different hierarchies between the mass scale and the jet and ultrasoft scales can
arise within a single spectrum when the kinematic variables are varied
within their allowed kinematic ranges. 

We stress that the mass mode method we present here is not tied to separating
the problem into a treatment of massive gauge bosons and dispersion integrations.
Once the field theoretic setup is identified (where the use of massive gauge
bosons might certainly serve as a useful guideline) the approach can be formulated
right away in terms of massive quark modes. In general this can be even necessary
when jet observables are considered that do not depend in a universal way on the
invariant mass of the 
quark pair. However, in cases where the dispersion method can be applied the
results for the massive gauge bosons are universal and results for other
types of secondary massive particles (scalars, vector bosons) can be readily
obtained simply by using the respective version of the dispersion
relation. Moreover in many non-universal cases the computationally easier
dispersion method can be used to at least correctly determine ultra-violet (UV)
divergent and/or infrared (IR) singular terms, such that the remaining finite
terms might not need to be carried out with regulators.
For thrust this is the case concerning large angle soft radiation. 

An interesting technical aspect of computations with mass modes is that
  usual dimensional regularization is not capable of separating the mass-shell
  contributions coming from the collinear and soft sectors,  see e.g.\ Refs.~\cite{Chiu:2007yn,Chiu:2007dg,Chiu:2009yx}.  
This leads to unregularized divergences in the respective collinear and soft
diagrams which appear to be related to the fact that dimensional regularization
regularizes with respect to virtualities, whereas the mass-shell
  fluctuations from the collinear and soft mass modes
are separated only by a boost. 
As we demonstrate in this work, these singularities drop out of the
  properly defined matrix elements and matching coefficients, so that in principle no regularization scheme other than dimensional regularization is necessary.
Using the mass mode massive gauge bosons and the dispersion method allows for a 
simple computational scheme to deal with these singularities because
the singular terms can already be canceled at the level
of one-loop diagrams.
This can also lead to substantial technical simplifications in cases where the
secondary massive quarks enter an observable in a non-universal way. 
The singularities appearing in the mass-shell contributions of the
  collinear and soft mass modes, however, lead to large logarithmic terms in
  the mass-shell matching corrections that arise when the mass modes
  are integrated out. These logarithms can in general not be summed up within
  the renormalization evolution carried by the $\mu$-dependence within
  dimensional regularization. Recently, several approaches to handle the
  summation of these ''rapidity logarithms" have been discussed in the
  literature \cite{Becher:2010tm,Becher:2011dz,Chiu:2011qc,Chiu:2012ir}. Since the methods to treat the rapidity logarithms are
  orthogonal to the renormalization flow governed by the $\mu$-dependence,
  they can be readily implemented into our  mass mode formalism through a
  separate procedure. We adopt a
  method based on the ``rapidity renormalization group``~\cite{Chiu:2012ir},
  and using an analytic regulator~\cite{Becher:2011dz}. The outcome is a simple
  exponentiation of logarithms in the mass mode matching conditions. Our treatment represents a novel non-trivial application of
  summing rapidity logarithms worth to be studied by itself.

We note that our effective theory setup involving the mass mode gauge
bosons entails that they are distinct degrees of freedom and exist in parallel
to the common massless collinear and ultrasoft gluons. While we primarily give them the
role of placeholders for the mass mode collinear quarks (and the collinear
gluonic modes they can interact with) the calculations
we carry out in this work can be of direct interest by themselves and might well serve for
concrete applications 
for example related to the electroweak theory. To be specific the mass mode gauge bosons might be considered
within the context of an additional spontaneously broken SU(2) gauge theory
where all SU(2) gauge bosons get a common mass $M$ from a Higgs in the 
fundamental representation. 
Similar to Refs.~\cite{Jantzen:2006jv,Chiu:2007yn,Chiu:2007dg} we write the corresponding SU(2) group
theory factors using $C_F$, $C_A$ and $T_F$ and denote the gauge coupling as
$g$ with $\alpha_s=g^2/4\pi$. This agrees with our notation for the QCD
results and facilitates the interpretation of the results for the dispersion
integration used to implement the corrections due to the gluon splitting. 

As a final introductory remark we note that for the calculations of the secondary
massive quark effects all computations are performed in dimensional
regularization. Since in this paper we mostly focus on the conceptual field theory
and computational issues related to the massive gauge bosons
we present all results in the limit $d=4-2\epsilon \rightarrow 4$ to facilitate
the discussions. Details on the $d$-dimensional results together with the
corrections caused by the secondary massive quarks are treated in a
subsequent publication~\cite{Hoang:2013}.

\section{The massless factorization theorem} \label{sect:massless}

In this section we briefly review the known massless factorization
theorem for thrust to set up our notations and collect the perturbative QCD results at one
loop. The inclusion of 
the mass modes modifies the form of the massless factorization theorem, and we show in the next sections that the
various mass mode contributions to the factorization theorem
interpolate continuously to the respective massless contributions in the limit $M\to 0$.
The factorization theorem sums large logarithms in the terms of
  the thrust distribution that are singular in the dijet limit. The thrust distribution reads~\cite{Fleming:2007qr}  
\begin{align}
\frac{d\sigma}{d\tau}=& \, \sigma_0 Q^2 H_0(Q,\mu_H) U^{(1)}_{H}(Q,\mu_H,\mu)
\nn\\
&\times\int ds\int\hspace{-0.1cm}ds^\prime\, J_0(s^\prime,\mu_J)\,U^{(1)}_J(s-s^\prime,\mu,\mu_J)  \nn \\ 
& \times  \int d\ell \, S_0(Q\tau-\frac{s}{Q}-\ell,\mu_S) U^{(1)}_S(\ell,\mu,\mu_S),
\label{eq:diffsigma0}
\end{align}
where $\sigma_0$ denotes the total partonic $e^{+}e^{-}$ cross-section at
tree-level, $H_0$ is the hard current matching condition, $J_0$ the thrust jet
function and $S_0$ the thrust soft function. 
Large logarithms between the characteristic scales of each sector, $\mu_H$,
$\mu_J$ and $\mu_S$, and the final renormalization scale $\mu$ are summed by the
evolution factors $U^{(1)}_{H}$, $U^{(1)}_J$ and $U^{(1)}_S$ satisfying the renormalization group equations
\begin{align}
\label{eq:currentRGE_massless}
 \mu\frac{d}{d\mu}U^{(1)}_{H}(Q,\mu_H,\mu) &=\gamma_H(Q,\mu)U^{(1)}_{H}(Q,\mu_H,\mu) \, , \\[3mm]
\label{eq:jetRGE_massless}
 \mu\frac{d}{d\mu}U^{(1)}_{J}(s,\mu,\mu_J)&=\int ds' \gamma_J(s-s',\mu)U^{(1)}_{J}(s',\mu,\mu_J) \, , \\
\label{eq:softRGE_massless}
 \mu\frac{d}{d\mu}U^{(1)}_{S}(\ell,\mu,\mu_S)&=\int d\ell'
 \gamma_S(\ell-\ell',\mu)U^{(1)}_{S}(\ell',\mu,\mu_S) \, .
\end{align}
The superscript $(1)$ in the evolution factors as well as the subscript $0$ in
the matrix elements indicate an expression in the massless theory without mass modes.
The choice of $\mu$ is arbitrary, and the dependence on $\mu$ cancels exactly working at a particular
order. In this work we adopt the choice
$\mu=\mu_S$, such that the evolution factor $U_S^{(1)}(\ell,\mu_S,\mu_S)=\delta(\ell)$ and can be
dropped from Eq.~(\ref{eq:diffsigma0}). The fact that other choices  for $\mu$
can be implemented leads to interesting
consistency conditions between the renormalization group factors which have been
discussed in detail in Ref.~\cite{Fleming:2007qr}. We stress that 
beginning at $\mathcal{O}(\alpha_s^2)$ all ingredients of the factorization
theorem depend on the number of massless quarks $n_f$. Currently $H_0$, $J_0$
and the partonic contributions to $S_0$ are known up to
$\mathcal{O}(\alpha_s^2)$ 
\cite{Moch:2005id,Becher:2006qw,Hoang:2008fs,Kelley:2011ng,Hornig:2011iu} and the anomalous
dimensions up to $\mathcal{O}(\alpha_s^3)$ \cite{Moch:2005id,Moch:2004pa}, for $H_0$ even the $\mathcal{O}(\alpha_s^3)$ corrections are available~\cite{Heinrich:2009be,Lee:2010cga} (see also Refs.~\cite{Becher:2008cf,Abbate:2010xh} for more detailed information).

The hard function is $H_0(Q,\mu)=|C_0(Q,\mu)|^2$ with the massless Wilson coefficient $C_0(Q,\mu)$ from
matching SCET to QCD. At $\mathcal{O}(\alpha_s)$ it reads (with $Q^2 \equiv Q^2 +i0$)
\begin{align}
 C_{0}(Q,\mu)=
1+\frac{\alpha_s C_F}{4\pi}&\left\{-\ln^2\left(\frac{-Q^2}{\mu^2}\right)
+3 \,\ln\left(\frac{-Q^2}{\mu^2}\right)\right.\nn\\
&\left.-8+\frac{\pi^2}{6}\right\}\, .
\label{eq:hardmatch0}
\end{align}
The current divergences are canceled by the current renormalization factor
\be
Z_C (Q,\mu)=1-\frac{\alpha_s C_F}{4\pi}\left\{\frac{2}{\epsilon^2}+\frac{3}{\epsilon}-\frac{2}{\epsilon}\,\ln\left(\frac{-Q^2}{\mu^2}\right)\right\}\, ,
\label{eq:Zc}
\ee  
which yields for the leading order anomalous dimension
\begin{align}\label{eq:gamma_H}
 \gamma_H (Q,\mu) &= - Z_C^{-1}(Q,\mu) \, \mu \frac{d}{d\mu}\, Z_C(Q,\mu) \, + \,c.c. \nn\\&= \frac{\alpha_s}{4\pi} \left\{2\Gamma_0\,\ln\left(\frac{Q^2}{\mu^2}\right) +\gamma^{H}_0\right\} \,,
\end{align}
with $\Gamma_0=4C_F$ and $\gamma^{H}_0=-12C_F$ being the ${\cal O}(\alpha_s)$ coefficients of the cusp and non-cusp anomalous
dimensions. 

The thrust jet function $J_{0}(s,\mu)$ is obtained through a convolution of the
hemisphere jet functions \cite{Fleming:2007xt} for the two jet directions,
\begin{equation}
J_{0}(s,\mu)=\int ds' \, J_{0,n}(s',\mu) J_{0,\bar{n}}(s-s',\mu) \,.
\end{equation}
The jet functions $J_n$ and $J_{\bar{n}}$ are matrix elements of collinear
fields in SCET and describe the dynamics of the collinear degrees of freedom. The
$n-$collinear jet function is defined as 
\begin{align}
 J_n(Q r_n^+,\mu)\equiv 
\frac{-1}{4\pi N_c Q}&\mathrm{Im}\left[ i 
\int {d^4x\;e^{i r_n \cdot x}}\right.\nn\\
&\langle 0|
\mathrm{T}\{\bar{\chi}_{n,Q}(0)\slash{\bar{n}}\chi_n(x)
\}|0\rangle\bigg]\,, 
\label{eq:jetfunctiondefinition}
\end{align}
where the jet fields $\chi_n(x),{\chi}_{n,Q}(0)$ denote quark fields multiplied
by collinear Wilson lines and the invariant mass is $r_n^2 \simeq Q r_n^+$. All
color and spin indices are traced implicitly. For further details and the
definition for $J_{\bar{n}}(Q r_{\bar{n}}^-,\mu)$ we refer to
Ref.~\cite{Fleming:2007qr}.   

The renormalized expression for $J_0(s,\mu)$ at one loop order reads 
\begin{align}\label{eq:jet0}
 \mu^2 J_{0}(s,\mu^2)=\delta(\bar{s})+\frac{\alpha_s
   C_F}{4\pi}&\left\{\delta(\bar{s})\left(14-2\pi^2\right)-6\left[\frac{\theta(\bar{s})}{\bar{s}}\right]_{+}\right.
 \nn\\
&\left.+8\left[\frac{\theta(\bar{s})\, \ln \,{\bar{s}}}{\bar{s}}\right]_{+} \right\}
\end{align}
with $\bar{s}\equiv s/\mu^2$. The one-loop renormalization factor for the massless jet
function reads
\be
\mu^2Z_{J}(s,\mu)=\delta(\bar{s})+\frac{\alpha_s C_F}{4\pi}\left\{\delta(\bar{s})\left(\frac{8}{\epsilon^2}+\frac{6}{\epsilon}\right)-\frac{8}{\epsilon}\left[\frac{\theta(\bar{s})}{\bar{s}}\right]_+\right\} \, ,
\label{eq:ZJ}
\ee
which yields for the leading order anomalous dimension
\begin{align}\label{eq:gamma_J}
 \mu^2 \gamma_J (s,\mu)& = - \int ds' Z_J^{-1}(s-s',\mu) \, \mu \frac{d}{d\mu}\, Z_J (s',\mu)\nn\\& = \frac{\alpha_s }{4\pi}\left\{ - 4\Gamma_0  \left[\frac{\theta(\bar{s})}{\bar{s}}\right]_{+} + \gamma^J_0 \, \delta(\bar{s})\right\} \, .
\end{align}
with $\gamma^{J}_0=12C_F$.

The thrust soft function $S_{0}\left(\ell,\mu\right)$ describes ultrasoft radiation
between the two jets and is based on a hemisphere definition,  
\begin{align}
 S_{0}(\ell,\mu)\equiv\frac{1}{N_c} \sum_{X_s}\, 
&\delta(\ell-\bar{n}\cdot k_s^R-n\cdot k_s^L)\, 
\langle 0 \lvert\overline{Y}_{\bar{n}} Y_n(0)\lvert X_s \rangle \nn\\&\times
\langle X_s \lvert Y_n^{\dagger} \overline{Y}_{\bar{n}}^{\dagger} (0)\lvert 0\rangle \, ,
\end{align}
where $k_s^R$ ($k_s^L$) is the momentum of the soft final state $\lvert X_s
\rangle$ in the right (left) hemisphere and $Y_n(x)$, $\overline{Y}_{\bar{n}}(x)$
are ultrasoft Wilson lines, i.e.\
\ba
 Y_n(x)&\equiv& \overline{\textrm{P}} \,  
\textrm{exp}\left[-ig\int_{0}^{\infty} ds \, n\cdot A_{us}(ns+x)\right]\,,\nn\\ 
\overline{Y}_{\bar{n}}(x)&\equiv& \overline{\textrm{P}} \,  
\textrm{exp}\left[-ig\int_{0}^{\infty} ds \, \bar{n} \cdot \overline{A}_{us}(ns+x)\right] \, .
\ea
In the tail region where the ultrasoft scale is larger than the hadronic scale
$\Lambda_{\rm QCD}$ the thrust soft function factorizes into a perturbative
partonic part and a nonperturbative hadronic part. In the following we will
consider just the partonic soft function. The renormalized expression at
$\mathcal{O}(\alpha_s)$ is 
\begin{equation}\label{eq:soft0}
\mu \, S_{0}(\ell,\mu)=
\delta(\bar{\ell})+\frac{\alpha_s C_F}{4\pi}
\left\{\frac{\pi^2}{3}\delta(\bar{\ell})
-16\left[\frac{\theta(\bar{\ell})\ln \, 
\bar{\ell}}{\bar{\ell}}\right]_{+}\right\} \,
\end{equation}
with $\bar{\ell}\equiv \ell/\mu$. Its renormalization factor at one loop reads
\be
\mu \, Z_{S}(\ell)=\delta(\bar{\ell})-
\frac{\alpha_s C_F}{4\pi}\left\{\frac{4}{\epsilon^2}\delta(\bar{\ell})
-\frac{8}{\epsilon}\left[\frac{\theta(\bar{\ell})}{\bar{\ell}}\right]_+\right\} \, ,
\label{eq:ZS}
\ee
and yields for the leading logarithmic anomalous dimension
\begin{align}
 \mu \gamma_S (\ell,\mu)& = - \int d\ell' Z_S^{-1}(\ell-\ell',\mu) \, \mu
 \frac{d}{d\mu}\, Z_S (\ell',\mu)\nn\\
& = \frac{\alpha_s }{4\pi} \left\{4\Gamma_0 \left[\frac{\theta(\bar{\ell})}{\bar{\ell}}\right]_{+}\right\} \, .
\end{align}
The anomalous dimensions (to any loop order)
satisfy the consistency relation
\be
\gamma_H(Q,\mu)\delta(\bar{s})+\mu^2\gamma_J(s,\mu)+\frac{\mu^2}{Q}\gamma_S(s/Q,\mu)\, =\,0.
\ee
\section{The Mass Mode Setup} \label{sect:setup}

When mass modes are included, their intrinsic mass scale can enter the field theory setup in
addition to the hard, jet and ultrasoft scales. It is 
convenient to define the ratio $\lambda_M=M/Q$. There are different types of
mass modes, collinear type modes and soft modes \cite{Chiu:2009yx}. If
  kinematically allowed, the $n$- and $\bar{n}$-collinear and soft mass modes
  have the scaling and the virtualities of their massless counterparts, but in
  addition we have to separately account for their mass-shell fluctuations. These have the scaling $p^\mu_n \sim
Q(\lambda^2_M,1,\lambda_M)$ and $p^\mu_{\bar{n}}\sim
Q(1,\lambda^2_M,\lambda_M)$ for the $n$- and $\bar{n}$-collinear mass mode,
respectively, while soft mass modes have 
$p^\mu_s\sim Q(\lambda_M,\lambda_M,\lambda_M)$. It is the purpose of the mass
mode formalism to separate these fluctuations in a systematic way when necessary.
 
The collinear massive quark interactions are determined from a massive quark
collinear Lagrangian~\cite{Leibovich:2003jd} which is a straightforward
generalization of the massless collinear Lagrangian.  
The soft mass mode gauge bosons couple to collinear quarks through soft Wilson
lines~\cite{Fleming:2007qr,Fleming:2007xt} and the interactions of soft mass mode
quarks and gluons among themselves is given by usual QCD interactions. 
An important aspect is that the mass mode
collinear gauge bosons are defined with a soft mass mode-bin
subtraction~\cite{Chiu:2009yx} in order to avoid double counting and maintain
gauge invariance within each collinear sector.  

\begin{figure*}
 \begin{center}
  \subfigure[$\lambda_M>1>\lambda>\lambda^2$] {\epsfig{file=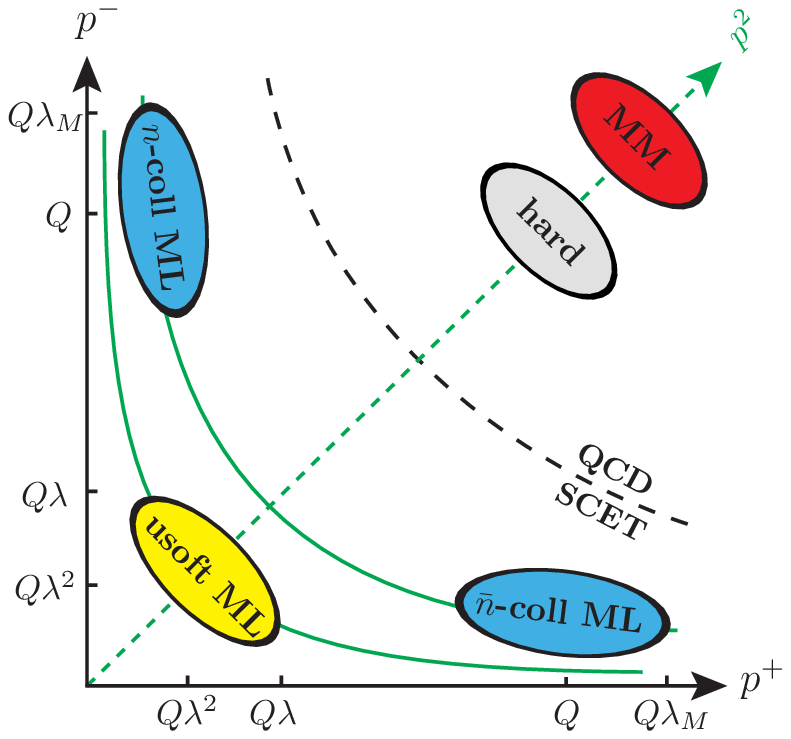,width=0.35\linewidth,clip=}}
  \subfigure[$1>\lambda_M>\lambda>\lambda^2$]{\epsfig{file=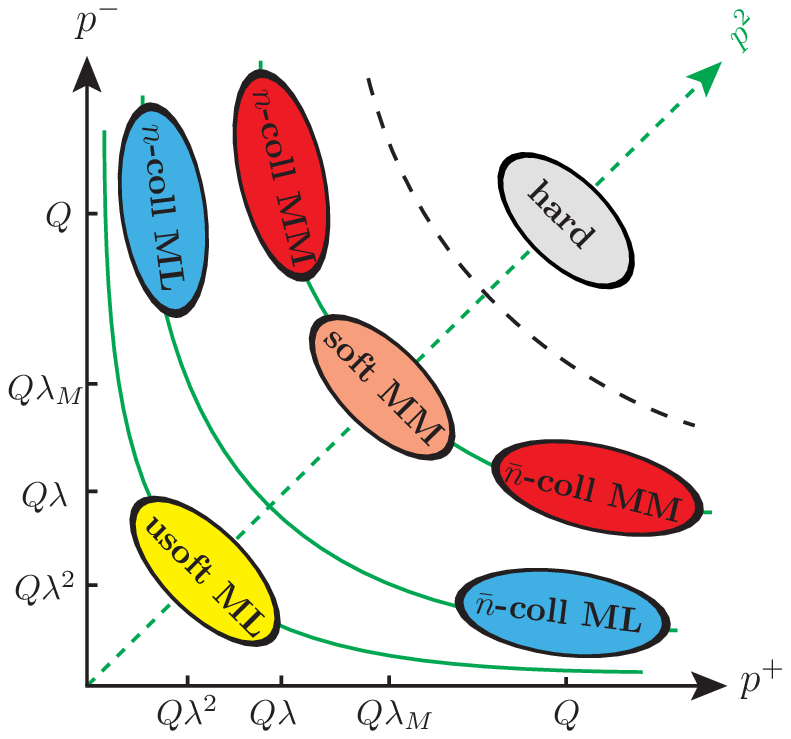,width=0.35\linewidth,clip=}}\\ 
    \subfigure[$1>\lambda>\lambda_M>\lambda^2$] {\epsfig{file=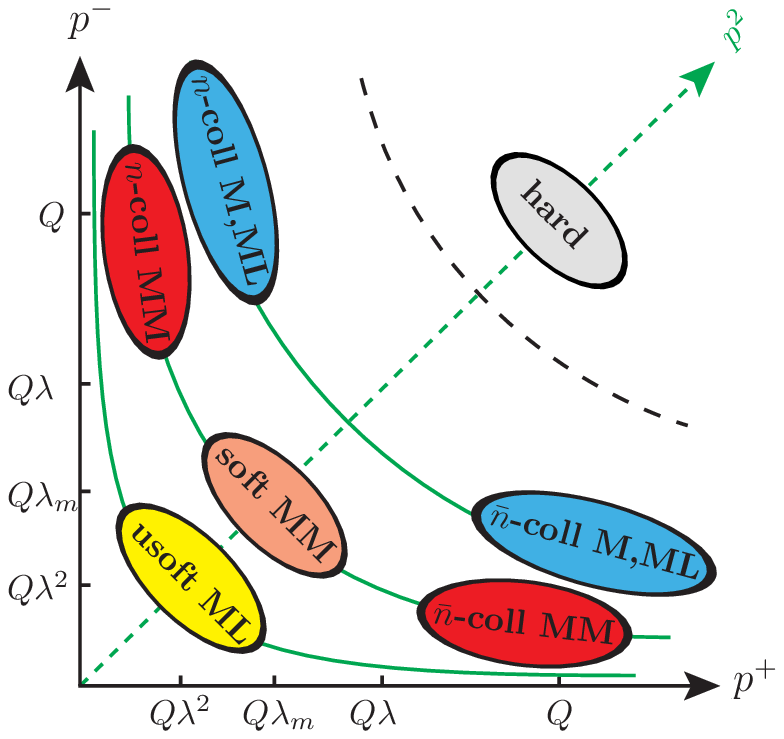,width=0.35\linewidth,clip=}}
  \subfigure[$1>\lambda>\lambda^2>\lambda_M$]{\epsfig{file=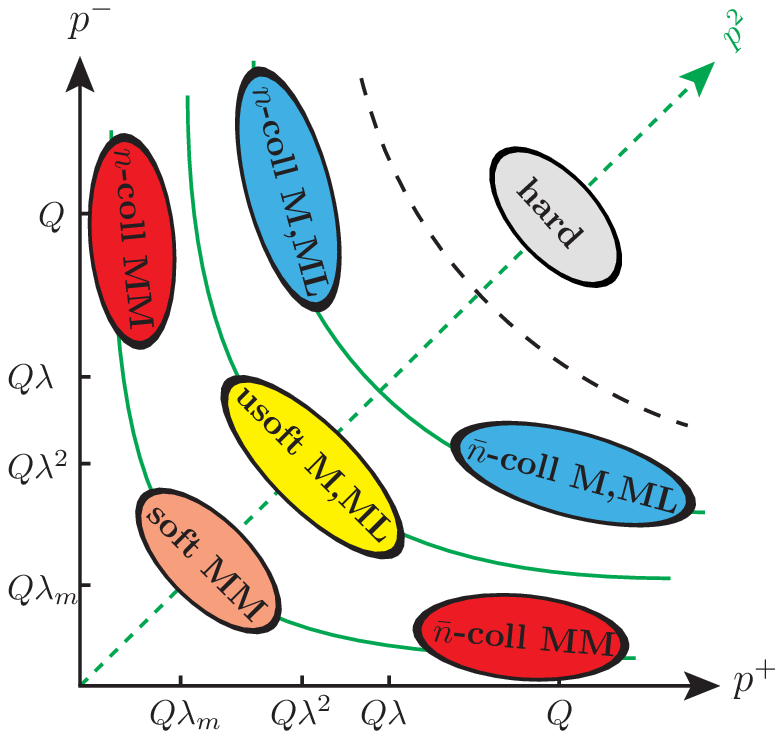,width=0.35\linewidth,clip=}} 
  \end{center}
  \caption{Localization of massless (ML) and massive (M) modes together with their mass-shell fluctuations (MM) in the  $p^{+}$-$p^{-}$phase space 
    according to their generic scaling for different
    hierarchies between $\lambda$ and $\lambda_M$. Modes with the same invariant
    mass are located on the same mass hyperbola. This is always the case for the collinear
    and soft mass-shell fluctuations.\label{fig:scaling}} 
\end{figure*}

\begin{figure*}
  \includegraphics[width=0.85\linewidth]{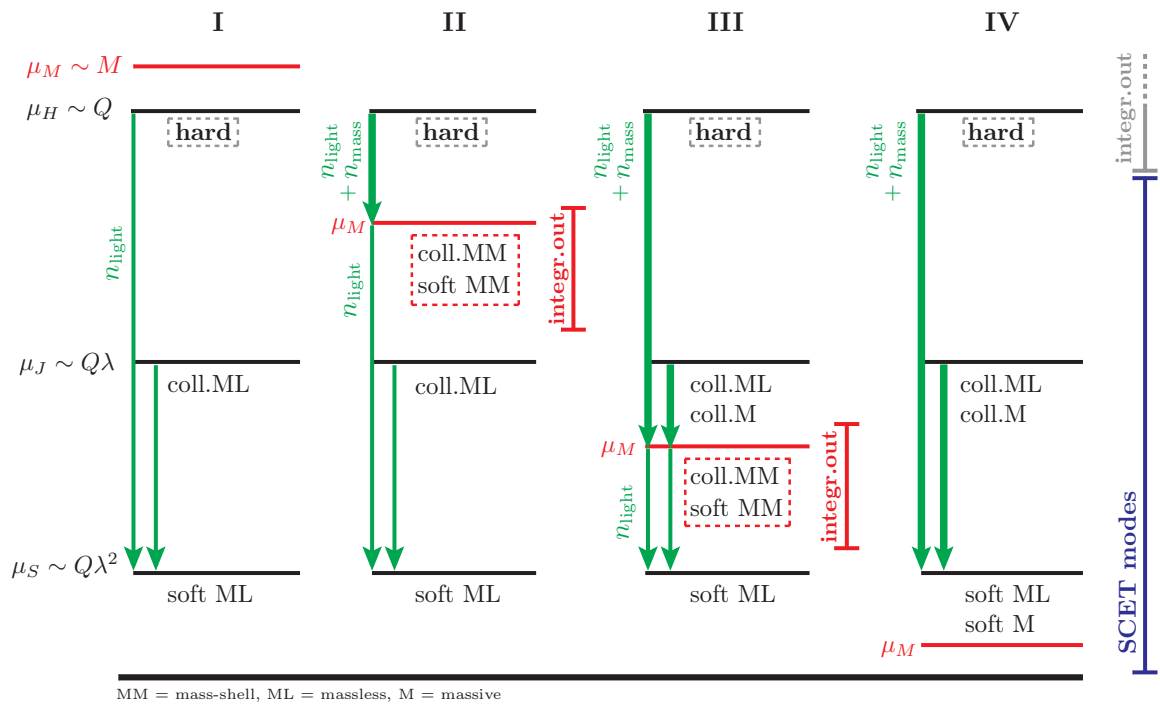}
  \caption{\label{fig:scenarios} 
The different scenarios depending on the
    hierarchy between the mass scale $M$ and the hard, jet and ultrasoft scale. MM
    indicates mass-shell scaling, ML the massless one. With M we denote modes
    that have a mass $M$ but scale as their massless counterparts. The
    renormalization group evolution is also shown in the top-down evolution from
    the hard scale $\mu_H$ down to $\mu=\mu_S$. When the mass scale is
    crossed the mass shell fluctuations are integrated out. This leads to 
    a matching condition and to a change in the evolution factor. In the case of
    secondary quark pairs production we evolve with $n_{\rm light}+n_{\rm mass}$ flavors above the mass scale and
    with $n_{\rm light}$ below ($n_{\rm mass}$ is the number of quark flavors with mass $M$).} 
\end{figure*}

The mass-shell fluctuations of the collinear and the soft mass modes are of the
same order with $\sqrt{p^2_{\bar{n}}}\sim\sqrt{p_n^2}\sim \sqrt{p_s^2}\sim M$ and thus just separated by boosts and rapidity.
They must be considered together with the collinear and
ultrasoft virtualities given by the kinematics related to the scales $Q\lambda$ and $Q\lambda^2$, respectively, with
$\lambda\sim \sqrt{\tau}$ for thrust in the tail region. Depending on the relative size of $Q$, $\lambda_M$ and $\lambda$, the mass modes
therefore have to be treated differently with respect to the massless collinear and ultrasoft
modes. This has an impact on the required matching and renormalization group
calculations and therefore also on the form of the resulting factorization
theorem.
The different scenarios can be ordered according to where the mass scale is
situated with respect to the jet and ultrasoft scales, see
Fig.~\ref{fig:scaling} and \ref{fig:scenarios} for graphical illustrations. Each
scaling situation corresponds in principle to a 
different EFT setup.

In the following we briefly summarize these different scenarios:\footnote{For the effective field theory setup we require $\lambda \ll 1$, but no large hierarchy concerning $\lambda_M$.}

\begin{enumerate}
\item[I)] $M>Q>Q\lambda>Q\lambda^2$ ($\lambda_M>1>\lambda>\lambda^2$): 
The mass $M$ is larger than the hard scale $Q$ and the mass modes are not
contained in SCET, but integrated out when SCET is matched to the full
theory. So the factorization theorem is the one for massless fermions up to
the hard current matching coefficient which acquires an additional
contribution due to the mass modes. The mass effects decouple for $\lambda_M \rightarrow \infty$. 

\item[II)] $Q>M>Q\lambda>Q\lambda^2$ ($1>\lambda_M>\lambda>\lambda^2$): 
In this scenario the mass $M$ is in between the hard scale and the jet
scale. The mass mode effects are virtual because the jet scale $Q\lambda$, the
typical invariant mass for real collinear particle radiation is below $M$. However,
the mass modes contribute to the 
matching condition at the scale $Q$ and the evolution of the current
for scales above $M$. The mass-shell fluctuations are integrated out at the scale $M$. The factorization theorem is
the one for the massless quarks concerning jet and soft function but receives
mass mode corrections in the matching and evolution of the production current. The
matching condition of the production current at the hard scale $Q$ has the correct
massless limit for $\lambda_M \rightarrow 0$, since all mass-shell
  contributions have been removed from it in the  matching procedure.

\item[III)] $Q>Q\lambda>M>Q\lambda^2$ ($1>\lambda>\lambda_M>\lambda^2$): 
The mass is below the jet scale. Therefore massless and massive collinear modes
both can fluctuate in the collinear sector yielding real and virtual contributions
to the jet function. The soft mass modes only arise through their virtual
effects since their interactions with collinear modes induce invariant masses
above the jet scale. They contribute to the current matching and evolution together with the
virtual mass-shell effects of the massive collinear modes exactly as in scenario
II. Analogously to the previous scenario the mass-shell fluctuations are integrated out
at the scale $\mu_M\sim M$ leading to an additional collinear mass mode
matching condition. The soft function agrees with the one of the
massless factorization theorem. 

\item[IV)] $Q>Q\lambda>Q\lambda^2>M$ ($1>\lambda>\lambda^2>\lambda_M$): When the
  mass is below the ultrasoft scale the mass-shell fluctuations of the mass
  modes merge into those of the massless quarks and gluons
    and in principle do not have to be separated. The collinear and soft fluctuations of
    the mass modes are treated at the same footing as those of the massless
    collinear and ultrasoft quarks and gluons. The soft mass modes adopt the same
  invariant mass scaling as the massless soft degrees of freedom. However, they
  still have a mass and this must be taken 
  into account in the calculations. In this last scenario the mass-shell fluctuations are
  not integrated out and the factorization theorem has the form of 
  the massless factorization theorem with the difference that the hard, jet and soft functions contain additional mass-dependent corrections which merge into the massless result for $\lambda_M\to 0$.
\end{enumerate}
An important aspect
is that the results in each scenario account for the full mass dependence of
the singular contributions, which we do not further expand if the mass becomes
smaller than any of the other kinematic scales. This yields a continuous description valid for any scaling between the mass and the hard, jet or soft
scales and is important when scanning the thrust variable through the entire
allowed spectrum. At this point one might worry whether the factorization theorems are
subject to uncontrolled power corrections at the point where two scenarios are
patched together. We show below that this is not the case and that there is no
generic loss of precision where the transition between different scenarios is
carried out.

We stress that the notation, the formulation of the factorization theorems and
the organization of the RG evolution we employ is associated to the 
``top-down'' evolution, where the scale $\mu$ is equal to the ultrasoft scale $\mu_S$.
Thus only the current and the jet function evolution factors $U_{H}$ and
$U_J$, respectively, appear in the factorization theorem. This also affects the
interpretation and association of the mass-shell contributions which enter in the mass mode matching conditions discussed
below. In this RG setting the jet and the soft functions do not receive any
massive quark effects when the jet and ultrasoft scales, respectively, are below the
mass mode matching scale. Alternative ways to describe RG-evolution related to
different choices 
for $\mu$ are possible and have been discussed in
Ref.~\cite{Fleming:2007xt}. There are consistency conditions that relate the 
RG-evolution and the mass mode matching factors for the different choices
  of $\mu$. Since the mass $M$ appears as an
additional scale in our mass mode setup, there are even more possibilities to
set up RG-evolution. Therefore the consistency conditions become more
involved, as they also entail relations between 
mass mode matching conditions and renormalization group evolution with
different anomalous dimensions and varying flavor number. As an example, if
$\mu$ is set to the hard scale, the jet and soft functions have to be
evolved upwards to larger scales. In this RG setup the jet and soft functions can 
pick up virtual mass-shell corrections in case the mass threshold is
crossed. We briefly come back to this issue in Sec.~\ref{sect:softfunction}
and discuss
the impact of these scenarios in more detail in Ref.~\cite{Hoang:2013}.

In the subsections~\ref{sect:scenarioI}-\ref{sect:scenarioIV} we give a more
detailed description of the formalism and the 
results for each of the scenarios for the application to thrust. We also
address
why the transition between the various scenarios is continuous and why
uncontrolled power corrections or power counting breaking effects do not arise.  Overall
this is 
conceptually related to the fact that the different effective theories are
all matched
to the singular contributions computed with the
  complete mass dependence in the full theory in an independent way and should not be thought of as being a
sequence 
of effective theories matched to each other. 

\subsection{Scenario I: $M>Q>Q\lambda>Q\lambda^2$}\label{sect:scenarioI}

The massive gauge boson with vector coupling is integrated out at the hard scale $Q$, when SCET is matched
to the full theory, yielding just a modification of the hard matching coefficient. It can be combined with
the massless coefficient $C_0(Q,\mu)$ to give a new Wilson coefficient, which
reads at one loop order 
\begin{equation}
\mathcal{C}^{I}(Q,M,\mu)=C_{0}(Q,\mu)+\delta F_{m}(M/Q)\, .
\label{eq:hardcoeffI}
\end{equation}
The massive contribution $\delta F_{m}(M/Q)$ is computed from the full theory
current form 
factor with a subtraction at $(p+p^\prime)^2=0$ (i.e.\ in the on-shell scheme which ensures
decoupling) since the
mass modes do not contribute to 
the SCET renormalization group evolution. At one-loop order
$\delta F_{m}(M/Q)$ reads~\cite{Kniehl:1988id,Hoang:1995fr,Hoang:1995ex}
\begin{align}
\delta F_{m}(x)=&\frac{\alpha_s C_F}{4\pi} \left\{ (1+ x^2)^2 \left[2 \, \Li_2
    (-x^2)- \ln^2(-x^2)\right. \right. \nn
\\  
& +\left.2 \, \ln(1+x^2) \,\ln(-x^2)-\frac{2\pi^2}{3}\right]
\nn\\
&- \left. (3+2x^2)\,\ln(-x^2) -2x^2 -\frac{7}{2} \right\}
\label{eq:hardQCDm}
\end{align}
with $x^2= M^2/(Q^2+i0)$. In the limit $M\to\infty$ the mass modes decouple,
so $\delta F_{m}(x) \rightarrow 0$ for $x \to \infty$. In the small mass limit $x
\rightarrow 0$ (which is not supposed to be taken in this scenario), on the
other hand, we obtain 
\begin{equation}
 \delta F_{m}(x) \stackrel{x\rightarrow 0}{\longrightarrow}  -
 \frac{\alpha_s C_F}{4\pi} \left\{\ln^2(-x^2)+ 3 \, \ln(-x^2) +\frac{2\pi^2}{3}
   +\frac{7}{2}\right\}\, ,
\label{eq:hardQCDmexpanded}
\end{equation}
which yields unresummed large logarithms. Thus in scenario~I the correct
massless limit cannot be obtained in the hard Wilson coefficient of
  Eq.~(\ref{eq:hardcoeffI}), since the latter still contains the mass-shell
  fluctuations of the mass modes. Since the 
SCET setup in this scenario is exactly the 
massless one, the factorization theorem adopts the form
\begin{align}
 \frac{d\sigma}{d\tau}&=Q^2\sigma_0\left|{\mathcal{C}^{I}(Q,M,\mu_H)}\right|^2U^{(1)}_{H}\left(Q,\mu_H,\mu_S\right)
 \int d\ell \int ds\nn\\
&  J_{0}(s,\mu_J) \, U^{(1)}_J(Q\ell-s,\mu_S,\mu_J) \, S_{0}\left(Q\tau-\ell,\mu_S\right)\,.
\label{eq:diffsigmaI}
\end{align}
In this scenario the jet and the soft functions only receive contributions from
the massless modes. The evolution factors $U^{(1)}_i$ employed in
Eq.~(\ref{eq:diffsigmaI}) are just the ones from the massless theory.  

\subsection{Scenario II: $Q>M>Q\lambda>Q\lambda^2$}\label{sect:scenarioII}

In this second scenario the mass modes enter SCET as long as their virtuality
is above $M$.\footnote{For $\lambda_M\ll 1$ this is the situation
    discussed in \cite{Chiu:2007yn,Chiu:2007dg,Chiu:2009yx}.} We have to
take the mass modes into account for the
matching to the full theory at the hard scale $\mu_H\sim Q$ and the RG-evolution above
$M$. Thus the RG-evolution between 
$\mu_H$ and  $\mu_M$, which affects the current evolution only, incorporates the
mass modes. When crossing the mass mode matching scale $\mu_M\sim M$ we
integrate out these massive 
modes. This leads to an additional current matching coefficient in comparison to the
massless factorization formula, which describes the mass-shell fluctuations. From there on everything proceeds as in the
massless case. The factorization theorem thus takes the form 
\begin{align}
 \frac{d\sigma}{d\tau}&=Q^2\sigma_0\left|{\mathcal{C}^{II}(Q,M,\mu_H)}\right|^2U^{(2)}_{H}\left(Q,\mu_H,\mu_M\right)\nn\\
&\times\left|{\mathcal{M}_{H}(Q,M,\mu_H,\mu_M)}\right|^2U^{(1)}_{H}\left(Q,\mu_M,\mu_S\right) \int d\ell \int ds\nn\\
&\, J_{0}(s,\mu_J) \, U^{(1)}_J(Q\ell-s,\mu_S,\mu_J) \, S_{0}\left(Q\tau-\ell,\mu_S\right)\,.
\label{eq:diffsigmaII}
\end{align}
In Eq.~(\ref{eq:diffsigmaII}) the superscript $(2)$ in the evolution kernel
$U^{(2)}_{H}\left(Q,\mu_H,\mu_M\right)$ indicates that the evolution is now
performed with the mass mode gauge bosons and the massless gluons. The hard matching
coefficient $\mathcal{C}^{II}(Q,M,\mu)$ acquires now a
subtractive contribution arising from the non-vanishing SCET
diagrams involving virtual collinear and soft mass mode gauge bosons. The
renormalized one-loop result of the corresponding sum of the virtual 
collinear and soft mass mode diagrams with on-shell external massless quarks
reads  
\begin{align}
&\delta F_{\textrm{eff}}(Q,M,\mu)=\frac{\alpha_s C_F}{4\pi}
\left\{2\,\ln\left(\frac{M^2}{\mu^2}\right)\ln\left(\frac{-Q^2}{\mu^2}\right)\right.
\nn\\
&\left.-\ln^2\left(\frac{M^2}{\mu^2}\right)-3\,\ln\left(\frac{M^2}{\mu^2}\right) -\frac{5\pi^2}{6}+\frac{9}{2} \right\}\,,
\label{eq:hardeffm}
\end{align}
in agreement with Ref.~\cite{Chiu:2009yx}. The UV-divergences are
mass-independent and agree with those from the massless diagrams. The
calculation is carried out explicitly in Sec.~\ref{sect:hardfunction} and does
not require any regularization in addition to dimensional regularization. 
For the calculation of the mass mode collinear diagrams one has to
account for soft mass mode bin 
subtractions\footnote{We acknowledge that the soft-bin
  subtractions can vanish in some regularization methods used for the
  rapidity divergences.}. The result for the 
total renormalized Wilson coefficient at the scale $\mu$ is given by 
\begin{equation}
\mathcal{C}^{II}(Q,M,\mu)=\mathcal{C}^I(Q,M,\mu)-\delta F_{\textrm{eff}}(Q,M,\mu) \, ,
\label{eq:hardcoeffmassive}
\end{equation}
with $\mathcal{C}^I(Q,M,\mu)$ being the hard current matching coefficient from
scenario~II in Eq.~(\ref{eq:hardcoeffI}).
For $M\rightarrow 0$ we now recover the correct massless limit, i.e. $\mathcal{C}^{II}(Q,M,\mu)
\stackrel{M\rightarrow 0}{\longrightarrow} 2 \, C_0(Q,\mu)$ with $C_0(Q,\mu)$
given in Eq.~(\ref{eq:hardmatch0}). So for $M\ll Q$ $\mathcal{C}^{II}$ is free of large logarithms at leading order in the $1/Q$ expansion for $\mu_H\sim Q$ .
The anomalous dimension of $\mathcal{C}^{II}(Q,M,\mu)$ is mass-independent and
just twice the one from the purely massless theory in Eq. (\ref{eq:gamma_H}),
so 
\begin{align}
\mu\frac{d}{d\mu}U^{(2)}_{H}(Q,\mu_H,\mu)=2\gamma_H(Q,\mu)U^{(2)}_{H}(Q,\mu_H,\mu)\,.
\end{align}
The hard current mass mode matching coefficient $\mathcal{M}_{H}(Q,M,\mu_H,\mu_M)$ is obtained by
integrating out the mass-shell fluctuations at the scale $\mu_M\sim
M$. The result reads at fixed order
\be
\left.\mathcal{M}_{H}(Q,M,\mu_M)\right|_{\rm{FO}}=1+ \delta F_{\textrm{eff}}(Q,M,\mu_M)\,.
\label{eq:matchingII}
\ee
and is obtained by matching to the ${\cal O}(\alpha_s)$  full theory result in
Eq.~(\ref{eq:fixedorder}) for $\tau<M^2/Q^2$ ($M>Q\lambda$), as explained in 
Sec.~\ref{sect:fulltheoryresults}. The matching correction in
Eq.~(\ref{eq:matchingII}) agrees exactly with 
the sum of the collinear and soft mass mode corrections involved already for the
hard matching condition of Eq.~(\ref{eq:hardcoeffmassive}) because the full
theory results relevant for scenarios~I and II are identical (being proportional
to $\delta(\tau)$). This is related
to the fact that the thresholds for mass mode production are located in
scenarios~III and IV so that the fixed-order full theory result (discussed in detail in
Sec.~\ref{sect:fulltheoryresults}) does not distinguish between scenarios~I and
II. In other words, the mass mode contributions $\delta 
F_{\textrm{eff}}(Q,M,\mu_M)$, which are subtracted from the scenario~I hard
matching coefficient to obtain the infrared-safe $\mathcal{C}^{II}(Q,M,\mu_H)$, 
are added back in the
mass mode matching coefficient when the mass modes are integrated out.
This is exactly the situation realized for hard coefficients in the ACOT
scheme~\cite{Aivazis:1993pi,Aivazis:1993kh}.
Since for $\mu_M=\mu_{H}\sim Q$, where the evolution factor
$U^{(2)}_{H}\left(Q,\mu_H,\mu_H\right)$ is unity, the very same term is swapped
between these two structures, the continuity of the spectrum at the transition
between scenarios I and II is ensured up to higher order $\alpha_s$-corrections
which do not contain large logs  and depend on the implementation of
the fixed order terms. This shows that the transition between scenarios~I and II
has to be carried out for $M\sim Q$.

Interestingly, the continuity of the transition from scenario~I to scenario~II
is related to the fact that the complete mass-dependence is
  incorporated in the hard current Wilson coefficient
  $\mathcal{C}^{II}$ of 
scenario~II. In the transition region $\mu_M\approx \mu_H$ the mass mode effective
field theory contributions in $\mathcal{C}^{II}(Q,M,\mu_H)$ and in the mass mode matching
coefficient $\mathcal{M}_{H}(Q,M,\mu_M)$ are proportional to 
$\delta F_{\textrm{eff}}(Q,M,\mu_H)-\delta F_{\textrm{eff}}(Q,M,\mu_M)\sim (\mu_H-\mu_M)/Q \ll 1$, so they vanish in the transition region. We note that the mass mode effective theory diagrams are  implemented at leading order in the $1/Q$ expansion, so that power counting breaking terms do not arise. These features will remain true at any order of perturbation theory in
$\alpha_s$.

Notice that for $\mu_M \ll \mu_H$ the mass mode matching coefficient
$\mathcal{M}_{H}$ contains a large logarithmic term $\sim
\Gamma_0\ln{(M^2/\mu^2_M)}\ln{(-Q^2/M^2)}$, which can be traced back to the
rapidity divergences contained in the collinear and soft mass mode diagrams
\cite{Chiu:2012ir}. For $\mu_M = M$ this logarithm is resummed by the
evolution factor $U^{(2)}_H$, but not any more for a generic choice $\mu_M
\sim  M$.\footnote{This feature holds generically just for the one-loop case. At higher orders one encounters a single logarithm which cannot be resummed in this way \cite{Jantzen:2006jv,Chiu:2007dg}.}  In Refs. \cite{Becher:2011dz,Chiu:2012ir,Chiu:2011qc} several approaches to sum these rapidity logarithms have been discussed. The outcome is that the large logarithmic term is simply exponentiated, and that the various  approaches correspond to scheme choices  that differ only  with respect to higher order terms that are not logarithmically enhanced. In Sec.~\ref{sect:hardfunction} we describe the 
calculation of the exponentiation formula following Refs.~\cite{Becher:2011dz,Chiu:2012ir,Chiu:2011qc} concerning the regularization of the rapidity divergences and the formulation of the resulting evolution equation. Expressing the limits of integration of the rapidity evolution equation in terms of the natural scaling properties of $\mu_H \sim Q$ and $\mu_M \sim M$ and expanding out all terms that do not involve a large logarithm,  the 
mass mode matching coefficient adopts the form

\begin{align}
&\mathcal{M}_H(Q,M,\mu_H,\mu_M)=\exp{\left[\frac{\alpha_s}{4\pi} \, \frac{\Gamma_0}{2} \, \ln{\left(\frac{M^2}{\mu^2_M}\right)}\ln{\left(\frac{\mu^2_H}{\mu^2_M}\right)}\right]}\nn
\\
&\times\left(1+\frac{\alpha_sC_F}{4\pi}\left\{2\,\ln\left(\frac{M^2}{\mu_M^2}\right)\ln\left(\frac{-Q^2}{\mu_H^2}\right)-\ln^2\left(\frac{M^2}{\mu_M^2}\right) \right.\right.\nn\\
&\left.\left.-3\,\ln\left(\frac{M^2}{\mu_M^2}\right)+\frac{9}{2}-\frac{5}{6}\pi^2\right\}\right) \, .
\label{eq:matchingII_RRG}
\end{align}
We note that the mass mode matching coefficient acquires a dependence on
the hard renormalization scale $\mu_H$ due to the summation of the large logarithms.

\subsection{Scenario III: $Q>Q\lambda>M>Q\lambda^2$}\label{sect:scenarioIII}
The mass is between the jet and the ultrasoft scale. As indicated in
Fig.~\ref{fig:scenarios}, massive and massless collinear modes can both
  fluctuate in the
same collinear sector yielding real and virtual contributions with typical invariant
mass of order $Q\lambda$. Thus massive and
massless collinear modes both contribute to the jet function. Between the jet
scale $\mu_J$ and the mass mode matching scale $\mu_M$ the jet function evolves
together with massive and massless collinear modes, and at $\mu_M$ the
mass-shell fluctuations are integrated out leading to an additional collinear matching 
coefficient which involves distributions. Concerning the hard current, its evolution
and matching with respect to
the mass mode matching scale $\mu_M$ agrees exactly with the one of 
scenario~II. Below the mass mode matching scale the current and jet
evolution as well as the soft function agree with the ones obtained in the well
known purely massless SCET setup. The form of the 
factorization theorem now reads
\begin{widetext}
\begin{align}
 \frac{d\sigma}{d\tau}=&Q^2\sigma_0\left|{\mathcal{C}^{II}(Q,M,\mu_H)}\right|^2U^{(2)}_{H}\left(Q,\mu_H,\mu_M\right)\left|{\mathcal{M}_{H}(Q,M,\mu_H,\mu_M)}\right|^2U^{(1)}_{H}\left(Q,\mu_M,\mu_S\right)  \nonumber\\  
&\times\int d\ell  \int ds \int ds' \int ds'' \; J_{0+m}(s,M,\mu_J) \, U^{(2)}_J(s'-s,\mu_M,\mu_J)\, \mathcal{M}_J(s''-s',M,\mu_J,\mu_M)\nonumber\\
&\times\,  U^{(1)}_J(Q\ell-s'',\mu_S,\mu_M)\,S_{0}\left(Q\tau-\ell,\mu_S\right)\, .
\label{eq:diffsigmaIII}
\end{align}
\end{widetext}
The matching coefficients $\mathcal{C}^{II}(Q,M,\mu)$ and
$\mathcal{M}_{H}(Q,M,\mu_H,\mu_M)$ are the same as in scenario~II, 
see Eqs.~(\ref{eq:hardcoeffmassive}) and (\ref{eq:matchingII_RRG}). 

The jet function $J_{0+m}(s,M,\mu)$ contains additional real and
  virtual contributions coming
from the collinear mass modes (including soft mass mode bin subtractions) and
has the form
\ba
J_{0+m}(s,M,\mu)&=&J_{0}(s,\mu) + \delta J^{\rm{virt}}_{m}(s,M,\mu) 
\nn\\[2mm] 
&& +  \delta J^{\rm{real}}_{m}(s,M)\,.
\label{eq:jetmassive}
\ea
For the
  massive collinear Feynman rules we used the counting $s\sim M^2$ to account
  for the full mass-dependence in the collinear sectors.
The term $\delta J^{\rm{virt}}_{m}(s,M,\mu)$ contains only
distributions and corresponds to virtual corrections. Its 
renormalized expression at one-loop order reads
($\bar s = s/\mu^2$)
\begin{align}
\label{eq:jetmassivevirt}
 &\mu^2 \delta J^{\rm{virt}}_{m}(s,M,\mu)= \frac{\alpha_s C_F}{4\pi} \left\{
 \delta(\bar{s})\left[-4\,\ln^2\left(\frac{M^2}{\mu^2}\right)\right.\right.
\\
&\left.\left.-6\,\ln\left(\frac{M^2}{\mu^2}\right)+9-2\pi^2\right] + \left[\frac{\theta(\bar{s})}{\bar{s}}\right]_{+}8 \,
  \ln\left(\frac{M^2}{\mu^2}\right)\right\} 
\,. \nn
\end{align}
The term $\delta J^{\rm{real}}_{m}(s,M)$ in Eq.~(\ref{eq:jetmassivereal}) contributes
only when the jet invariant mass is above the threshold $M$ and thus corresponds
to real production of the massive gauge boson.
At ${\cal O}(\alpha_s)$ it has the form
\begin{align}
% \mu^2 
& \delta J^{\rm{real}}_{m}(s,M)= 
%\mu^2 
\frac{\alpha_s
   C_F}{4\pi}\,
\theta(s-M^2)\,
\left\{\frac{2(M^2-s)(3s+M^2)}{s^3}\right.\nn\\
&\left.\hspace{2cm}+\frac{8}{s}\,\ln{\left(\frac{s}{M^2}\right)}\right\} \, . 
 \label{eq:jetmassivereal}
\end{align}
Due to its physical character it is UV-finite and does not
contain any explicit logarithmic $\mu$-dependence.
Furthermore, $\delta J^{\rm{real}}_{m}(s,M)$ is zero
at the threshold, so that no discontinuity arises due to real radiation. This
feature remains true also in the ${\cal O}(\alpha_s^2)$ massive quark correction
as the dispersion integration does not offset this property. For
$M\rightarrow 0$ the correct massless limit is reached by 
combining the real radiation pieces and the virtual contributions properly into 
distributions yielding 
 $J_{0+m}(s,M,\mu) \stackrel{M\rightarrow 0}{\longrightarrow} 2 \,
 J_{0}(s,\mu)$. So for $M^2 \ll s $ the jet function  $J_{0+m}(s,M,\mu)$ is
 free of large logarithms at leading order in the $1/Q$ expansion for $\mu^2 = \mu^2_J \sim s$. Moreover, using the massive collinear Feynman rules does not involve any power counting breaking as we strictly kept terms that are leading in the $1/Q$ expansion.
Details on the calculation of 
$\delta J^{\rm{virt}}_{m}$ and $\delta J^{\rm{real}}_{m}$ are presented in
Sec.~\ref{sect:jetfunction}. 
We also note that the massive collinear diagrams contain rapidity
  divergences. Including the
soft mass mode bin subtraction these divergences cancel and the loop integrals are successfully
regularized by dimensional regularization, so that there is no need to
introduce an additional regulator. 
The anomalous dimension coming from the virtual mass mode contribution $\delta
J^{\rm{virt}}_{m}$ is mass independent and exactly coincides with the one of the 
massless jet function in Eq.~(\ref{eq:gamma_J}). This yields
\be 
\mu \frac{d}{d\mu} U^{(2)}_J(s,\mu,\mu_0)=\int{d s'2 \gamma_J(s-s',\mu) U^{(2)}_J(s',\mu,\mu_0)}
\ee
for the RG-equation of the jet function above the mass scale $M$, where the RHS
is just twice the result obtained for the case when the mass modes do not contribute, see Eq.~(\ref{eq:jetRGE_massless}).
The RG-equation for the current evolution factor $U_{H}^{(2)}$ above $M$ remains
unchanged with respect to scenario~II. 

The collinear mass mode matching coefficient $\mathcal{M}_J(s,M,\mu_J,\mu_M)$ is
obtained when the mass-shell fluctuations are integrated out from the jet function in its
evolution down to the ultrasoft scale. At one-loop the fixed-order result reads 
\be
\left.\mathcal{M}_J(s,M,\mu_M)\right|_{\rm{FO}}=\delta(s)-\delta J^{\rm{virt}}_{m}(s,M,\mu_M) \,,
\label{eq:MJet}
\ee
and encodes just the virtual contributions contained in the mass mode jet
function. Equation~(\ref{eq:MJet}) is obtained by matching to the full theory
result in the $\tau$ region relevant for scenario~III, $M^2/Q^2<\tau<M/Q$
($Q\lambda>M>Q\lambda^2$), as 
shown in Eq.~(\ref{eq:scetscenarioIII}).
In the collinear mass mode matching coefficient only terms corresponding to the
distributive virtual part of the mass mode jet 
function appear as required by consistency, since matching coefficients do not
contain kinematic terms involving thresholds.

We note that for the thrust distribution the transition from the factorization theorem
of scenario~II to the one of scenario~III should be made for values of thrust for which
the jet scale is somewhat below the threshold of collinear massive real
radiation, i.e. $\tau \leq M^2/Q^2$, so that the threshold is properly
  accounted for through the analytic form of $\delta J^{\rm{real}}_{m}(s,M,\mu_M)$. 

Compared to the factorization
theorem of scenario~II we find that 
apart from the differences in the RG-evolution the distributive collinear mass mode
contributions in $J^{\rm{virt}}_{m}(s,M,\mu)$ are just swapped between the jet
function and the collinear mass mode matching coefficient. Thus for
$\mu_M=\mu_J$ where $U^{(2)}_J(t-s'',\mu_M,\mu_J)=\delta(t-s'')$
the factorization theorem of
scenario~III in Eq.~(\ref{eq:diffsigmaIII}) agrees with the one of
scenario~II in Eq.~(\ref{eq:diffsigmaII}) (up to higher order $\alpha_s$-corrections
which do not contain any large logarithms and depend on the implementation of
the fixed order terms). This ensures the continuity of the
transition in the theoretical descriptions between scenarios~II and III and is
in close analogy to the continuity we already discussed in the transition
between scenarios I and II. 
Since no hierarchy is assumed otherwise between the mass and the jet scale, the full $M$ dependence of the most singular terms is accounted for and scenario~III has the same
generic precision in the $1/Q$ expansion for all allowed $M$ values including the transition region
to scenario~II.   

Similar to the mass mode matching coefficient of the
  current $\mathcal{M}_H$, also $\mathcal{M}_J$ contains a large logarithm $\sim \Gamma_0\ln{(M^2/\mu^2_M)}\ln{(s/M^2)}$ for $\mu_M \ll \mu_J$,
  which is manifest when using the normalized jet invariant mass variable
  $\tilde{s} \equiv s/\mu^2_J$
\begin{align}
 \mu_J^2& \left.\mathcal{M}_{J}(s,M,\mu_M) \right|_{\rm{FO}}= \delta{(\tilde{s})}+\frac{\alpha_s C_F}{4\pi} \left\{\delta(\tilde{s})\left[4\,\ln^2\left(\frac{M^2}{\mu_M^2}\right)\right.\right. \nn \\
&\hspace{0.5cm} -\left. 8 \,\ln\left(\frac{M^2}{\mu_M^2}\right)\ln\left(\frac{\mu^2_J}{\mu_M^2}\right)+6 \,\ln\left(\frac{M^2}{\mu_M^2}\right)-9+2\pi^2\right]\nn\\
&\hspace{0.5cm}\left.+\left[\frac{\theta(\tilde{s})}{\tilde{s}}\right]_{+} \left[-8 \,\ln\left(\frac{M^2}{\mu_M^2}\right)\right] \right\}\, . \label{eq:MJet_0}
\end{align}
For the choice $\mu_M = M$ all logarithmic terms are resummed by the
evolution factor $U^{(2)}_J$, but not any more for a generic choice $\mu_M
\sim  M$. We can sum the logarithms applying the method already used for the current mass mode matching coefficient. In Sec.~\ref{sect:jetfunction} we
describe the calculation which again leads to an exponentiation. 
Expressing the limits of integration of the rapidity evolution equation in
terms of the natural scaling properties of $\mu^2_J\sim s$ and $\mu_M \sim M$
and expanding out all terms not involving any large logarithms the collinear
mass mode matching coefficient adopts the form
\begin{align}\label{eq:MJet_RRG}
&\mu^2_J\mathcal{M}_J(s,M,\mu_J,\mu_M)=\exp{\left[-\frac{\alpha_s}{4\pi} \,2\Gamma_0 \,\ln{\left(\frac{M^2}{\mu^2_M}\right)}\ln{\left(\frac{\mu^2_J}{\mu^2_M}\right)}\right]}\nn
\\
&\times\left(\delta{(\tilde{s})}+\frac{\alpha_sC_F}{4\pi}\left\{\delta{(\tilde{s})}\left[4\,\ln^2{\left(\frac{M^2}{\mu^2_M}\right)}+6\,\ln{\left(\frac{M^2}{\mu^2_M}\right)}-9\right.\right.\right.\nn\\
&\hspace{0.5cm}\left.+\left.2\pi^2\bigg]+\left[\frac{\theta{(\tilde{s})}}{\tilde{s}}\right]_+ \left[-8 \,\ln{\left(\frac{M^2}{\mu_M^2}\right)}\right]\right\}\right) \, .
\end{align}
In analogy to the current case discussed above, the collinear mass mode
  matching coefficient acquires a dependence on the jet scale $\mu_J$ due to the summation of the large logarithms.
  
\subsection{Scenario IV $Q>Q\lambda>Q\lambda^2>M$}\label{sect:scenarioIV}

If the mass scale is below the ultrasoft scale the massive modes are not integrated
out at all because we have $\mu \sim \mu_S >M$ and the invariant mass of the
ultrasoft fluctuations of order $Q\lambda^2$ exceeds the mass mode scale $M$. In
this scenario the collinear and soft mass modes fluctuate in the collinear and
soft sector, respectively, together with the corresponding massless
modes. Thus the mass scale $M$ becomes merely a parameter within the theory
that is not relevant any more for the separation of modes.  
Therefore the collinear as well as the soft mass modes are
contained in the massless collinear and ultrasoft sectors, respectively. In
this scenario the
massless and the massive degrees of freedom both contribute to all renormalization
group evolution factors as well as to the jet and the soft functions, and the
factorization theorem does not contain any mass mode matching condition.
Since we want to maintain the full mass-dependence of the singular terms
we keep the $M$ dependence in all contributions of the factorization theorem. The factorization theorem of scenario~IV has the form
\begin{align}
 \frac{d\sigma}{d\tau}=&Q^2\sigma_0\left|{\mathcal{C}^{II}(Q,M,\mu_H)}\right|^2U^{(2)}_{H}\left(Q,\mu_H,\mu_S\right)\nn\\ &\times\int d\ell  \int ds\,  J_{0+m}(s,M,\mu_J)\,U^{(2)}_J(Q\ell-s,\mu_S,\mu_J)\nn\\&\times \, S_{0+m}\left(Q\tau-\ell,M,\mu_S\right)\,.
\label{eq:diffsigmaIV}
\end{align}
Compared to the factorization theorems of scenarios~II and III
the hard current and jet mass mode matching coefficients have disappeared.  
Instead there is a new term in the thrust soft function due to the soft mass mode gauge bosons, 
\begin{align}
S_{0+m}\left(\ell,M,\mu\right)=&S_{0}(\ell,\mu)+\delta
S^{\rm{virt}}_{m}(\ell,M,\mu)\nn\\&+ \delta S^{\rm{real}}_{m}(\ell,M) \,.
\label{eq:softmassive}
\end{align}
For the
  massive soft Feynman rules we used the counting $\ell\sim M$ to account
  for the full mass-dependence in the ultrasoft sector.
In analogy to the mass mode contributions to the jet function the term 
$\delta S^{\rm{virt}}_{m}(\ell,M,\mu)$ contains only distributions and
corresponds to virtual corrections. At one-loop order it reads 
($\bar{\ell}\equiv \ell/\mu$)
\begin{align}
 \label{eq:softmassivevirt}
 \mu \, \delta S^{\rm{virt}}_{m}(\ell,M,\mu) = \, & \frac{\alpha_s
   C_F}{4\pi}\left\{\delta(\bar{\ell}) \left[2
   \,\ln^2\left(\frac{M^2}{\mu^2}\right) +\frac{\pi^2}{3}\right]\right.
\nn\\
&\left.-8\,\ln\left(\frac{M^2}{\mu^2}\right)
  \left[\frac{\theta(\bar{\ell})}{\bar{\ell}}\right]_{+}\right\} \,.
\end{align}
We note that in the calculation of $\delta S^{\rm{virt}}_{m}(\ell,M,\mu)
  $ we find again rapidity divergences which, however, cancel between the
  contributions coming  from the two hemispheres.
 Thus the sum  is successfully regularized by dimensional regularization. The term $\delta S^{\rm{real}}_{m}(\ell,M)$ describes real
radiation of the soft massive gauge bosons and has the form
\begin{align}
\label{eq:softmassivereal}
\delta S^{\rm{real}}_{m}(\ell,M) = \,& \frac{\alpha_s C_F}{4\pi}
\theta(\ell-M)
\left\{-\frac{8}{\ell} \,\ln \left(\frac{\ell^2}{M^2}\right) \right\} 
\, .
\end{align}
It is UV-finite and does not contain any explicit $\mu$-dependence.  We also find
that $\delta S^{\rm{real}}_{m}(\ell,M)$ is zero at threshold, which remains
also true for the corrections involving secondary massive quark pair
production. For $M\rightarrow 0$ the correct massless limit is reached by combining the real
radiation pieces and the virtual contributions properly into distributions
yielding  $S_{0+m}(\ell,M,\mu) \stackrel{M\rightarrow 0}{\longrightarrow} 2 \,
S_{0}(\ell,\mu)$. Thus the mass mode contributions to the soft function
are free of large logarithms at leading order in the $1/Q$ expansion for
$\mu_S\sim\ell$. Mass singularitiesdo not occur in any ingredient of the factorization theorem at this point.
The anomalous dimension arising from the massive contribution
to the soft function is mass independent and 
coincides with the one of the massless contribution. This yields 
\be 
\mu \frac{d}{d\mu} U^{(2)}_S(\ell,\mu,\mu_0)=\int{{\rm d} \ell'\,2
  \gamma_S(\ell-\ell',\mu) U^{(2)}_S(\ell',\mu,\mu_0)}
\ee
for the RG-evolution for the soft function above the mass scale $M$,
where the RHS is just twice the result obtained when the mass modes do not
contribute, see Eq.~(\ref{eq:softRGE_massless}).
However, since we have adopted the scale setting $\mu=\mu_S$, no evolution
of the soft function has to be accounted for. 
Notice that the evolution factors  $U_{H}^{(2)}$ and $U_{J}^{(2)}$ in
Eq.~(\ref{eq:diffsigmaIV}) remain unchanged with respect to scenarios~II and III.

The transition between scenarios~III and IV is carried out for $M\sim Q\tau$.
It is continuous in close analogy to the
transitions between the previous scenarios discussed above. However, the check
is somewhat more involved here as it is based on the consistency relation 
\begin{align}
2 \, \textnormal{Re}&\left[{\delta F_{\rm
      eff}(Q,M,\mu)}\right]\delta(\tau)-Q^2\delta
J^{\rm{virt}}_{m}(Q^2\tau,M,\mu)\nn\\
& -\,Q \delta S^{\rm{virt}}_{m}(Q\tau,M,\mu)=0 \, 
\label{eq:smoothnessconstr}
\end{align}
of the different virtual mass mode contributions. The relation can be verified
by explicit analytic calculations, see the corresponding 
results given in Eqs.~(\ref{eq:hardeffm}), (\ref{eq:jetmassivevirt}) and
(\ref{eq:softmassivevirt}). We note that in practice the transition from
scenario~III to IV should be made for values of thrust for which the soft scale is
somewhat below the threshold of soft massive real radiation, i.e. $\tau \leq
M/Q$ to ensure that the threshold is properly accounted for through the
analytic form of $\delta{S}^{{\rm real}}$.
Since besides $M\leq Q\tau$ no hierarchy is assumed  between the mass and the soft scale, the full $M$ dependence of the most singular terms is accounted for and scenario~IV has the same
generic precision in the $1/Q$ expansion for all allowed $M$ values including the transition region
to scenario~III.   

The consistency
relation~(\ref{eq:smoothnessconstr}) expresses the fact that virtual mass mode contributions to the
hard current, the jet function and the soft function can be reshuffled among each other.
It is part of an extended set of more general
consistency relations that is associated to the freedom in fixing the
renormalization scale $\mu$ to in principle any scale below $Q$.
The way how the virtual mass mode contributions
appear and are interpreted depends on the choice of $\mu$. 
For example if we set $\mu=\mu_H$ (``bottom-up'' approach, where  the soft and
the jet functions have to be evolved to $\mu_H$), there are mass mode ``matching''
contributions for the soft and the jet function, but none associated to the
hard current contribution. In this case, the soft mass mode matching contains
the term $\delta S^{\rm{virt}}_{m}$ and for $\mu_M > \mu_S$ the corresponding large logarithm has to be resummed by exponentiation analogously to Eqs.~(\ref{eq:matchingII_RRG}),~(\ref{eq:MJet_RRG}). We describe this calculation in Sec.~\ref{sect:softfunction}.
A more detailed discussion on the consistency relations will be given in
Ref.~\cite{Hoang:2013}. 

\section{Fixed order full theory result}
\label{sect:fulltheoryresults}
The mass mode matching coefficients given in the scenarios outlined in
Sec.~\ref{sect:setup} are derived by matching the differential cross
section in the different effective theory scenarios to the fixed order result
(\ref{eq:fixedorder}) obtained in the full theory. 
In this section we present the ${\cal O}(\alpha_s)$ full
theory result by calculating the diagrams with massive gauge bosons shown in  
Fig.~\ref{fig:QCDdiag1}. We also discuss the various expansions needed
to identify the singular terms, which are relevant
for the
matching calculations in the different field theory scenarios. Since the full
theory result has non-trivial threshold contributions related to the mass scale, 
these expansions deserve a separate discussion.

\subsection{General Result}
\label{subsect:fulltheoryresults}
\begin{figure}[t]
 \centering
 \includegraphics[width=\linewidth]{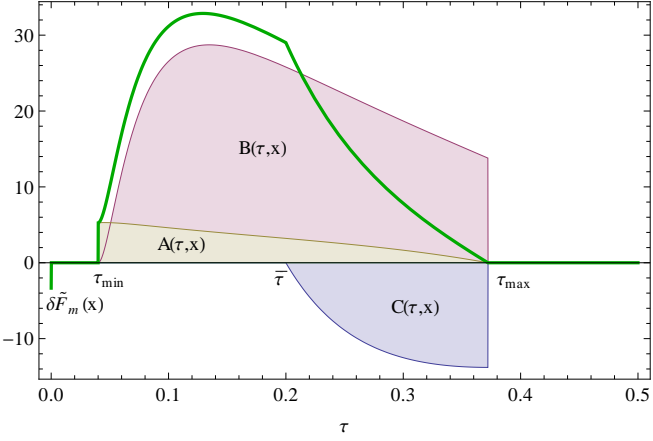}
 \caption{The three contributions $A(\tau,x)$, $B(\tau,x)$ and
   $C(\tau,x)$ to the ${\cal O}(\alpha_s)$ full theory thrust distribution for
   the radiation of a 
   massive gauge boson for $x=M/Q=0.2$. The sum of all terms, given by the
   thick green line, is positive. The vertical green line at $\tau=0$ indicates
   the  virtual corrections proportional to $\delta(\tau)$.\label{QCDthrust}} 
\end{figure}

The virtual contributions are known~\cite{Kniehl:1988id,Hoang:1995fr,Hoang:1995ex} and yield $\delta
F_m(x)$ in Eq.~(\ref{eq:hardQCDm}) using on-shell renormalization for the
massless external quarks. 
Gauge invariance together with the on-shell normalization
condition generate automatically $\delta F_m=0$ at $(p+p^{\prime})^2=0$ \cite{Hoang:1995fr}. What
is left to be done, is the calculation of the diagram for real radiation 
of the massive gauge bosons and the integration 
\begin{figure}[H]
 \centering
 \includegraphics[width=0.9\linewidth]{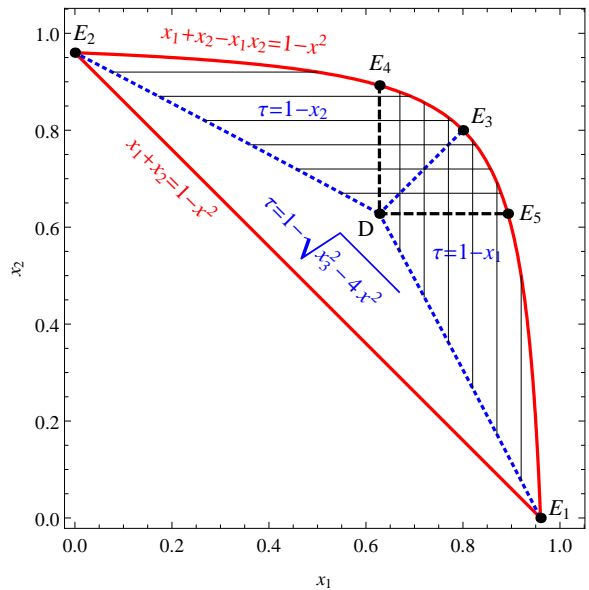}
 \caption{The phase space for the real radiation of a massive gauge boson in
   terms of the variables $x_1$ and $x_2$ for $x=M/Q=0.2$. The kinematic 
   constraints yield the allowed area within the red, continuous lines. The
   blue dotted lines bound the areas with the same thrust relation to $x_1$ and
   $x_2$ and are given by $x_1=x_2$, $2x_1(2-x_2)=(2-x_2)^2-4x^2$ and
   $2x_2(2-x_1)=(2-x_1)^2-4x^2$. $D$ (located at 
   $x_1=x_2=\sqrt{x_3^2-4x^2}=1-\tau_{\rm max}$) is the point of maximal thrust, $\rm E_1$ and
   $\rm E_2$ have minimal thrust configurations. At $E_3$ the produced gluon is
   at rest with $\tau=\bar{\tau}$. The triangle $\rm E_1 D E_2$ corresponds to
   $\tau=1-\sqrt{x_3^2-4x^2}$ and generates the function
   $A(\tau,x)$. $B(\tau,x)$ is obtained by increasing the areas of the segments
   $\rm E_1 E_3 D$ ($\rm E_3 E_2 D$) to $\rm E_1 E_4 D$ ($\rm E_5 E_2 D$) and
   integrating with $\tau=1-x_1$ ($\tau=1-x_2$). Subtracting $\rm E_5 E_4 D$
   with the corresponding opposite thrust prescription gives the soft function
   piece $C(\tau,x)$.\label{FO_massive}}  
\end{figure}
over the three particle phase space. 
It
is convenient to introduce the usual energy fraction variables, 
\be
x_1=\frac{2p_0}{Q},\quad x_2=\frac{2p_0^\prime}{Q},\quad
x_3=\frac{2q_0}{Q},\quad x=\frac{M}{Q}\,,
\ee
where $x_1+x_2+x_3=2$. Here $p$ ($p'$) is the quark (antiquark)
momentum, $M$ and $q$ are the gauge boson mass and momentum, respectively, see Fig.~\ref{fig:QCDdiag1}. The
double differential cross section  in $d=4$ dimensions then reads
\begin{align}
\frac{1}{\sigma_0}&\frac{d\sigma^{\textnormal{real}}}{dx_1\, dx_2}=
\frac{\alpha_s C_F}{2\pi}\frac{1}{(1-x_1)(1-x_2)}\bigg[\, x_1^2 \nn\\[2mm]
&-\,x^2\left( \frac{1-3x_2+2x_1 x_2}{1-x_1} \right)+x^4 +(x_1
  \leftrightarrow x_2)\,\bigg]
\,.
\label{eq:doublesigma}
\end{align}
The relation of the thrust variable $\tau$ to $x_{1,2,3}$ reads
\begin{align}
\tau &\equiv 1- \max_{\hat{\mathbf{t}}}\frac{\sum_i |\hat{\mathbf{t}} \cdot
  \vec{p}_i|}{Q} \nn\\
 & =1-\textnormal{max}\left(x_1,x_2,\sqrt{x_3^2-4x^2}\right)\, .
\label{eq:thrust}
\end{align}
The resulting thrust distribution (including also the virtual contributions)
displays structures to thresholds associated to collinear and soft gauge boson
radiation. It has the form 
\begin{align}
\frac{1}{\sigma_0}\frac{d\sigma}{d\tau} = & \frac{\alpha_s
  C_F}{4\pi}\bigg\{\delta \tilde{F}_m(x) \,
\delta(\tau)\nn\\
&+\theta(\tau-\tau_{\textnormal{min}})\,\theta(\tau_{\textnormal{max}}-\tau)\,
\Big(A(\tau,x)+B(\tau,x)\Big)\nn\\
& +\theta(\tau-\bar{\tau})\,\theta(\tau_{\textnormal{max}}-\tau)\, C(\tau,x)\bigg\} \, ,
\label{eq:totalsigma}
\end{align}
where the minimal and maximal thrust values and the intermediate threshold for
the real radiation contributions are given by ($x=M/Q$)  
\begin{align}
\tau_{\textnormal{min}} \equiv  x^2\,,\quad \tau_{\textnormal{max}} \equiv \frac{1}{3}\left(-1+2\sqrt{1+3x^2}\right)\, , \quad \bar{\tau}  \equiv  x \, .
\end{align}
The real corrections associated to the threshold at $\tau_{\rm
  min}=M^2/Q^2$ correspond to collinear radiation, and those associated to the
threshold at $\bar\tau=M/Q$ correspond to soft radiation. As illustrated
in Fig.~\ref{FO_massive}, we have distributed the contributions 
from the various phase space regions such that collinear and soft terms both
extend to the endpoint at $\tau_{\rm max}$. This facilitates the
identification of the singular terms necessary for the matching calculations
carried out in the SCET framework since in this way they are compatible with the
contributions coming from the jet and the soft functions. We emphasize
  that this setup appears to be the only practical choice to achieve a
  separation of collinear and soft contributions compatible with singular
  terms that can be analytically defined for the whole kinematic $\tau$-range
  and for all possible values of $x=M/Q$. Note that in Eq.~(\ref{eq:totalsigma}) we have factored out an overall loop
factor and redefined $\alpha_s C_F/2\pi \delta
\tilde{F}_m(x)\equiv\textrm{Re}[\delta {F}_m(x)]$ with $\delta {F}_m(x)$ given
in Eq.~(\ref{eq:hardQCDm}). The functions $A(\tau,x)$, $B(\tau,x)$ and $C(\tau,x)$ are obtained by
integrating $d\sigma^{\textnormal{real}}/dx_1\, dx_2$ in the phase-space
regions as described in the caption of Fig.~\ref{FO_massive}. The analytic results read 
\begin{widetext}
\begin{align}
 \label{eq:A}
 A(\tau,x) =&  \frac{(\hat{w}-2\tau)(1-\tau)}{\tau\hat{w}(\hat{w}-\tau)}\left[1-\hat{w}^2-2\tau(1+2\hat{w})+5\tau^2 \right] \nn \\
 & +\frac{1-\tau}{2\hat{w}^2} \left[9-12\hat{w}+14\hat{w}^2 -4\hat{w}^3+\hat{w}^4+4\tau(3-2\hat{w}+\hat{w}^2)- 2\tau^2 (1-\hat{w})^2-4\tau^3+\tau^4\right] \ln \left( \frac{\hat{w}-\tau}{\tau} \right) \, , \\
 B(\tau,x) =&  \frac{(1+\hat{w}-3\tau)(1+\tau-\hat{w})}{8\tau^3 (\hat{w}-\tau)} \left[\hat{w}(1-\hat{w}^2)-\tau(1+14\hat{w}-5\hat{w}^2+4\hat{w}^3) +\tau^2(7-4\hat{w})(2-w)+\tau^3(3+4\hat{w})-4 \tau^4 \right]\nn \\
 &  + \frac{1}{2\tau}\left[(3+\hat{w}^2)^2-2\tau^2 (1+\hat{w}^2)+\tau^4\right] \ln \left( \frac{4\tau(\hat{w}-\tau)}{\hat{w}^2-(1-\tau)^2} \right) \, ,  \label{eq:B} \\
 C(\tau,x) = & -\frac{1}{8 \tau^3} \left[1-\hat{w}^2-2\tau+5 \tau^2\right] \left[1-\hat{w}^2-2\tau(7+2\hat{w}^2)-11 \tau^2+4 \tau^3\right] \nn \\
 & -  \frac{1}{2\tau} \left[\left(3+\hat{w}^2\right)^2-2\tau^2(1+\hat{w}^2)+ \tau^4\right] \ln \left(\frac{4 \tau^2}{w^2-(-1+\tau)^2}\right) \, , \label{eq:C}
\end{align}
\end{widetext}
with $\hat{w}  \equiv\sqrt{(1-\tau)^2+4 x^2}$. Our result agrees with the one given in Ref.~\cite{Gardi:2000yh}. The result for the fixed-order thrust
distribution in arbitrary units is displayed for $x=0.2$ in
Fig.~\ref{QCDthrust}.
As a cross-check, from Eq.~(\ref{eq:totalsigma}) we obtain the correct massless
expression for $x\rightarrow 0$ \cite{Ellis:1980wv,Abbate:2010xh},
\begin{align}
\frac{1}{\sigma_0}&\frac{d\sigma^{\textnormal{real}}}{d\tau}
\stackrel{M\rightarrow 0}{\longrightarrow}
\frac{\alpha_sC_F}{4\pi}\left\{2\delta(\tau)\left(-1+\frac{\pi^2}{3}\right)-6\left[\frac{\theta(\tau)}{\tau}\right]_+\right.\nn\\
&-8\left[\frac{\theta(\tau) \, \ln
    \,\tau}{\tau}\right]_++\frac{2}{1-\tau}\left[6+3\tau-9\tau^2+(2-4\tau)\,
  \ln \, \tau\right.\nn\\
&\left.\left.+\left(\frac{4}{\tau}-6+6\tau\right)\ln (1-2\tau)\right]\right\}
\, ,
\label{eq:massless}
\end{align}
valid for $0\le\tau\leq 1/3$. To obtain this result involving the
plus-distributions it is required to
exactly account for the analytic behavior of the collinear and soft thresholds when
taking the massless limit. 

\subsection{Expansions}
\label{subsect:fulltheoryexpansions}

The massless fixed order thrust distribution shown in Eq.~(\ref{eq:massless}) only
depends on $\tau$, and the expansion parameter relevant for approaching the dijet
limit is simply $\lambda\sim \sqrt{\tau}\ll 1$. There is only one single
threshold located at $\tau=0$. Thus in the massless limit we can 
identify the fixed-order singular pieces that are being summed in
the factorization theorem and that can be used for the matching
calculations within SCET by simply performing an expansion in small $\tau$. The
collinear and soft contributions are not separated.
In the presence of the massive gauge bosons the collinear and soft thresholds
are separated due to their different mass-dependence: collinear radiation
arises for $\tau\geq \tau_{\rm min}= M^2/Q^2$ and soft radiation arises
for $\tau \geq \bar\tau =M/Q$. Thus the identification of the singular
terms -- which shall be defined for the whole kinematic $\tau$-range and include the full
mass-dependence -- needs to account for the location of the thresholds. In this
subsection we discuss the required expansions for the full theory results and
show a simple example how it is applied in a matching calculation.  

The $\delta(\tau)$ term is left unexpanded since we aim at determining the full
mass-dependence of the singular contributions. The term proportional to
$\theta(\tau-x^2)$ contains the collinear contributions. Here the
jet invariant mass $~Q\sqrt{\tau}$ and the mass $M$ have to be considered at the
same footing. Since the
mass threshold is treated exactly, we expand for $\tau\sim x^2\ll
1$. Here only the function $B$ contains singular terms as it originates from the
collinear regions in the $x_1$-$x_2$~phase space where $x_1\sim 1$ or $x_2\sim 1$: 
\begin{align}
B(\tau,x) \stackrel{\tau \sim x^2}{\longrightarrow} &
 \, \frac{2}{\tau^3}\,\bigg[\,(x^2-\tau)(3\tau+x^2)+4\tau^2 \ln
 \left(\frac{\tau}{x^2}\right)\,\bigg]
\nn\\
& + \,\mathcal{O}(\tau^0,x^0)
\,.
\end{align}
The function $B$ and all terms in the expansion vanish at $\tau=x^2$ giving a
continuous turn-on of the singular ${\cal O}(\tau^{-1})$ terms (with $\tau\sim
x^2$). On the other hand, the function $A(\tau,x)$ 
does not contain singular ${\cal O}(\tau^{-1})$ terms as it stems from
configurations which are neither collinear nor soft. Interestingly the function
$A$ is non-vanishing at $\tau=x^2$ and responsible for the step visible in
Fig.~\ref{QCDthrust} at the $\tau=x^2$. For $\tau\sim x^2\ll 1$ the expansion
for the function $A$ reads
\begin{align}
A(\tau,x)\stackrel{\tau \sim x^2}{\longrightarrow}  
 -\,4 \,\frac{\tau+x^2+\tau \ln \, \tau}{\tau} +\mathcal{O}(\tau,x^2)\, 
\end{align}
showing that it contributes at ${\cal O}(\tau^0\sim 1)$.
Although the contribution from function $A$ exceeds the singular contribution
from $B$ numerically at $\tau\approx x^2$, it only contains non-singular terms which are not described
in the factorization theorems. It remains to discuss the term proportional to
$\theta(\tau-x)$, which contains the soft contributions. Here the soft scale
$Q\tau$ and the mass $M$ have to be considered at the same footing. Since the
soft mass threshold is treated 
exactly we expand in $\tau\sim x \ll 1$. For the function $C$ this yields singular
${\cal O}(\tau^{-1})$ terms which read 
\begin{align}
C(\tau,x)\stackrel{\tau \sim x}{\longrightarrow}
 -\,\frac{16}{\tau}\, \ln\left(\frac{\tau}{x}\right) + {\cal O}(\tau^0,x^0)
\,.
\end{align}
Similar to $B$, the singular $\mathcal{O}(1/\tau)$ terms turn on continuously at $\tau=x$. 

Assembling all singular pieces to be considered for the matching calculations we obtain 
%\begin{widetext}
\begin{align}
&\frac{1}{\sigma_0}\left. \frac{d\sigma^{\textnormal{full th.}}}{d\tau}\right|_{\rm FO}\, =  \,
\delta(\tau)+ 2 \,\mbox{Re}\left[\delta F_m (M/Q)\right]\,\delta(\tau) 
\nn \\&
+\frac{\alpha_s
  C_F}{4\pi}\left\{\,\theta(Q^2\tau-M^2)\left[\frac{2(M^2-Q^2\tau)(3Q^2\tau+M^2)}{Q^4\tau^3}\right.\right.
\nn\\
& \quad \left.\left.+\frac{8}{\tau}\,\ln{\left(\frac{Q^2\tau}{M^2}\right)}\right]
%\right. \nn\\ &\left.
-\theta(Q\tau-M)\frac{8}{\tau}\,\ln{\left(\frac{Q^2\tau^2}{M^2}\right)}\right\}\, .
\label{eq:fixedorder}
\end{align}
%\end{widetext}
We emphasize that the factorization theorems for {\it all} EFT scenarios
  discussed in Sect.~\ref{sect:setup} account for the full-mass dependence
  displayed in Eq.~(\ref{eq:fixedorder}).
The terms corresponding to the collinear and soft thresholds yield functions
which match exactly to the real radiation effective theory contributions 
from collinear (see $\delta J^{\rm real}_m$ in
Eq.~(\ref{eq:jetmassivereal})) and soft (see $\delta S^{\rm real}_m$ in
Eq.~(\ref{eq:softmassivereal})) mass modes, respectively.
We note that from Eq.~(\ref{eq:fixedorder}) the singular terms in the massless
limit (first three terms on the RHS of Eq.~(\ref{eq:massless})) can be
recovered, requiring the proper combination of the terms into 
plus-distributions and $\delta$-functions in $\tau$. 

An interesting issue we would like to mention is that the parametric precision of the
expansions leading to the singular contributions in Eq.~(\ref{eq:fixedorder}) is
not uniform and depends on $\tau$. Above the collinear threshold
($\tau>\tau_{\rm min}$) the expansion is valid up to higher order power
corrections of order $M^2/Q^2$. On the other hand, above the soft threshold
($\tau>\bar\tau$) the expansion is valid up to higher order power corrections of
order $M/Q$. This is an intrinsic property and also inherited 
to the factorization theorems we derived in the various scenarios. 
So, from a strict power counting point of view, for $M/Q\ll1$ the soft sector would need to be
treated with the first order power corrections included to reach the same
parametric precision as the collinear sector. We stress, however, that this does
not affect in any way the consistency of treating only the singular
collinear and soft mass mode contribution in the effective theory description.
In practice this might make the treatment of subleading power corrections more
important for the soft mass modes than for the collinear ones. 

As an example for the matching procedure from the underlying theory to SCET we
now derive 
the one-loop collinear mass mode matching contribution $\mathcal{M}^{(1)}_J$ in
scenario~III where $M^2/Q^2<\tau<M/Q$. Expanded in fixed order (i.e.\ setting
$\mu=\mu_H=\mu_J=\mu_M=\mu_S$), the mass mode contributions to the
factorization theorem for scenario~III in Eq.~(\ref{eq:diffsigmaIII}) read
\begin{align}
\frac{1}{\sigma_0}&\left.\frac{d\sigma^{\textnormal{SCET}}}{d\tau}\right|^{\textnormal{III}}_{\rm{FO}}=
\, \delta(\tau)+2 \,\textnormal{Re}\left[\delta
  F_m(M/Q)\right]\delta(\tau)\nn\\
&+Q^2 \left[\delta
  J^{\rm virt}_{m}(Q^2\tau,M,\mu)+\mathcal{M}_J^{(1)}(Q^2\tau,M,\mu) \right] \nn \\[2mm]
  & +Q^2\delta J^{\rm real}_m(Q^2\tau,M) \, .  
\label{eq:scetscenarioIII}
\end{align}
To determine $\mathcal{M}^{(1)}_J$ we have to compare this result to the
expanded full theory result of Eq.~(\ref{eq:fixedorder}) for
$M^2/Q^2<\tau<M/Q$. This yields the fixed order condition
\begin{equation}
\left.\mathcal{M}^{(1)}_J (Q^2\tau,M,\mu)\right|_{\rm FO}=-\delta J^{\rm virt}_m (Q^2\tau,M,\mu)\,.
\end{equation}
canceling exactly the distributive pieces of the massive jet function as described in
Sec.~\ref{sect:scenarioIII}. All other mass mode matching coefficients are
derived in a completely analogous way.

\section{Calculations with Mass Modes}
\label{sect:eftresults}

In this section we describe the effective theory calculations for the mass mode
contributions to the infrared-safe hard Wilson coefficient, the jet and the
soft functions and the mass-mode matching coefficients entering the factorization theorems in
Sec.~\ref{sect:setup}. For the regularization of all Feynman diagrams we use
dimensional regularization. The integration measure in light-cone coordinates
reads 
\be
\label{eq:lightconetransformation}
\frac{{\rm d}^d k}{(2\pi)^d}\quad\longrightarrow\quad 
\frac{1}{2} \frac{{\rm d}k^+}{2\pi}\frac{{\rm d}k^-}{2\pi}
\frac{2^{3-d}\pi^{1-d/2}}{\Gamma\left(\frac{d-2}{2}\right)}
\,{\rm d}k_\perp\;k_\perp^{d-3}
\ee
with $k_\perp\equiv|\vec{k}_\perp|$ for integrals not depending on the angles
between the transverse momenta. 

An important technical point is that we encounter ``rapidity'' divergences in single diagrams
which are not regularized by dimensional regularization. In the calculations
of the mass-mode contributions to the hard current Wilson coefficient
  ($\mu_H\sim Q$), the jet function ($\mu_J\sim Q\sqrt{\tau}$) and the soft
  function ($\mu_S\sim Q\tau$), we 
do not need to employ an additional regulator (see e.g.\
Refs.~\cite{Chiu:2007yn,Chiu:2007dg,Chiu:2009yx}). Here, these 
divergences turn out to cancel among the proper set of
diagrams and, if needed, the soft mass mode bin subtractions, and
  they do not result in large logarithms. We stress that for $M$ above
  the soft scale the collinear and the
soft mass modes also contain mass-shell contributions with the same
  typical invariant mass of order $M$. The mass-shell fluctuations
contribute to the mass mode matching coefficients that arise when the mass
modes are integrated out
(at $\mu_M\sim M$). For the mass mode matching coefficients the "rapidity"
divergences mentioned before also cancel, but they leave large logarithmic
terms that in general cannot be summed within the $\mu$-evolution formalism based on
dimensional regularization~\cite{Becher:2010tm,Chiu:2011qc,Chiu:2012ir}. For the mass mode matching
coefficient for the hard production current these logarithms are known to
exponentiate~\cite{Chiu:2011qc,Chiu:2012ir}. In Sec.~\ref{sect:hardfunction} we reproduce this result using the
regularization and evolution method from Refs.~\cite{Becher:2011dz} and \cite{Chiu:2012ir}, respectively, and we demonstrate in Secs.~\ref{sect:jetfunction} and~\ref{sect:softfunction},
using the same method, that analogous exponentiations arise for the collinear and soft mass mode
matching coefficients, respectively. 

We also note that the soft mass mode bin subtractions that can arise for
calculations of collinear mass mode corrections are somewhat different from the zero-bin
subtraction~\cite{Manohar:2006nz,Idilbi:2007yi,Idilbi:2007ff} for massless
collinear diagrams, which can be 
related to vanishing large collinear label momenta. 
%As shown in
%\cite{Chiu:2009yx} one can alternatively separate the 
%phase space into momentum regions. 
We note that we have checked explicitly that the contributions coming
from longitudinal polarizations of the massive gauge bosons vanish due to
gauge invariance. To achieve this for the calculations involving collinear mass
modes the inclusion of soft mass mode bin subtractions turned out to be
crucial. So the soft mass mode bin subtractions are essential to isolate
gauge-invariant structures and to achieve factorization, as has also been
stressed in Ref.~\cite{Chiu:2009yx}.
In the following we  present all calculations in Feynman gauge.

\subsection{Vertex Corrections and Current Matching}
\label{sect:hardfunction}

\subsubsection{Hard matching coefficient}
\begin{figure}
 \centering
 \includegraphics[width=\linewidth]{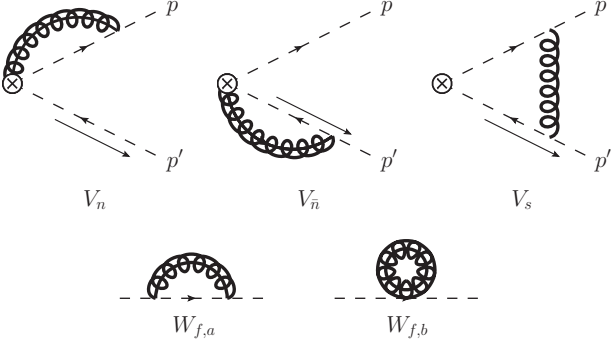}
 \caption{Non-vanishing EFT diagrams for the computation of the hard matching coefficient,
   soft mass mode bin subtractions are implied for the collinear
   diagrams. \label{fig:hardfunction_diagrams}}  
\end{figure}

In this section we discuss the calculation of the collinear and soft mass mode 
contributions to the effective theory vertex corrections feeding into the
matching for the hard current Wilson coefficient at the matching scale
$\mu_H\sim Q$ shown in 
Eq.~(\ref{eq:hardeffm}). The result is valid for all scenarios where the
mass scale $M$ is below the hard scale $Q$ (see Eqs.~(\ref{eq:diffsigmaII}),
(\ref{eq:diffsigmaIII}) and (\ref{eq:diffsigmaIV})). This can be understood from
the fact that the way how collinear and soft mass modes are distributed with
respect to the massless modes  below the
hard scale $Q$ does not affect the hard contribution itself.\footnote{
We carry out the calculations for external massless collinear quarks which are
on-shell. This leads to exactly the same calculations for all scenarios with
$M<Q$.}
We note that the same
result has been obtained before in Refs.~\cite{Chiu:2007yn,Chiu:2007dg,Chiu:2012ir} and~\cite{Chiu:2009yx} using an
additional analytic regulator and the $\Delta$-regulator, respectively. 
Here we show the calculation once more using only dimensional regularization
and the fact that all singularities cancel once soft mass mode bin subtractions
are included. We also clarify the role of the soft mass mode bin subtractions
for the collinear wave-function renormalization factors. The relevant collinear and 
soft diagrams with massive gauge bosons to be calculated are shown in
Fig.~\ref{fig:hardfunction_diagrams}, where $p$ ($p^\prime$) denotes the
momentum of the collinear massless external quark (antiquark).

Applying the SCET Feynman rules gives for the soft mass mode diagram
\begin{align}
V_s=& 2i g^2 \tilde{\mu}^{2\epsilon} C_F\int{\frac{d^dk}{(2\pi)^d}\frac{p^--k^-}{[(p-k)^2+i\se]}}\nn\\
&\times\frac{p^{\prime+}+k^+}{[(p^\prime+k)^2+i
    \se]}\frac{1}{k^2-M^2+i\se}\,\bar{\xi}_{n,p}\Gamma\xi_{\bar{n},p^\prime}
\,. 
\end{align}
Using $p=(0,Q,0)$ and $p'=(Q,0,0)$ and the soft scaling
$k^\mu\sim Q(\lambda_M,\lambda_M,\lambda_M)$ for the soft momentum we obtain
for $\lambda_M\ll 1$ 
\begin{align}
V_s&=2i g^2 \tilde{\mu}^{2\epsilon} C_F
\int{\frac{d^dk}{(2\pi)^d}\frac{1}{[-k^++i\se]}\frac{1}{[k^-+i\se]}}\nn\\
&\hspace{1cm}\times\frac{1}{[k^2-M^2+i\se]}\,\bar{\xi}_{n,p}\Gamma \xi_{\bar{n},p^\prime} \nn\\[2mm]
&\equiv 2i g^2\tilde{\mu}^{2\epsilon}C_F \, I_s \,\bar{\xi}_{n,p}\Gamma \xi_{\bar{n},p^\prime} \, .
\label{eq:defis}
\end{align}
The expression in Eq.~(\ref{eq:defis}) can as well be obtained directly from the
SCET Feynman rules upon field redefinition such that the soft mass mode gauge
bosons couple to the collinear massless quarks through mass mode Wilson
lines~\cite{Fleming:2007qr,Chiu:2009yx}.
The $n$-collinear diagram yields 
\begin{align}
V_n&=2i g^2 C_F\tilde{\mu}^{2\epsilon}
\int{\frac{d^dk}{(2\pi)^d}\frac{p^--k^-}{[(p-k)^2+i\se]}\frac{1}{[k^-+i\se]}}\nn\\
&\hspace{1cm}\times\frac{1}{[k^2-M^2+i\se]}
\,\bar{\xi}_{n,p}\Gamma\xi_{\bar{n},p^\prime}
\,,
\label{eq:Vn}
\end{align}
and using the form of the external momenta (and likewise the collinear scaling
$k^\mu\sim Q(\lambda_M^2,1,\lambda_M)$) we obtain 
\begin{align}
V_n&=2i g^2 C_F\tilde{\mu}^{2\epsilon} \int{\frac{d^dk}{(2\pi)^d}
\frac{Q-k^-}{[k^+(k^--Q)-k_\perp^2+i\se]}}
\nn\\
&\hspace{1cm}\times\frac{1}{[k^-+i\se][k^2-M^2+i\se]}
\,\bar{\xi}_{n,p}\Gamma\xi_{\bar{n},p^\prime} \nn \\[2mm]
&\equiv 2i g^2 C_F \tilde{\mu}^{2\epsilon}\,I_n
\,\bar{\xi}_{n,p}\Gamma\xi_{\bar{n},p^\prime} \, .
\label{eq:defin}
\end{align}
The $\bar{n}$-collinear diagram $V_{\bar{n}}$ is obtained in an analogous
manner and involves the function $I_{\bar{n}}$. It is obtained from the
$n$-collinear diagram by swapping the plus- with the 
minus-components as well as $p$ with $p^\prime$. Expanding the $n$-collinear
diagram of Eq.~(\ref{eq:Vn}) in the soft mass mode
regime $k\sim Q(\lambda_M,\lambda_M,\lambda_M)$ we obtain the 
soft mass mode bin subtraction contribution, 
\begin{align}
V_{n,0M}&=2i g^2 C_F\tilde{\mu}^{2\epsilon}
\int{\frac{d^dk}{(2\pi)^d}\frac{Q}{[-Qk^+ +i\se][k^-
    +i\se]}}  
\nn\\
&\hspace{1cm}\times\frac{1}{[k^2-M^2+i\se]}\,\bar{\xi}_{n,p}\Gamma\xi_{\bar{n},p^\prime} \nn \\
&\equiv 2i g^2 C_F \tilde{\mu}^{2\epsilon}I_{n,0M} 
\,\bar{\xi}_{n,p}\Gamma\xi_{\bar{n},p^\prime} \, ,
\end{align}
with the analogous expression for the $\bar{n}$-collinear sector, which we call
$I_{\bar{n},0M}$. Comparing with Eq.~(\ref{eq:defis}) we find the known relation
$I_{n,0M}=I_{\bar{n},0M}=I_{s}$ and the sum of all vertex diagrams including the
soft mass mode subtractions gives~\cite{Chiu:2009yx} 
\be
\delta F_{\rm{eff}} \sim (I_{n}-I_{n,0M})+(I_{\bar{n}}-I_{\bar{n},0M})+I_s= I_n+I_{\bar{n}}-I_s 
\label{eq:Isum}
\ee
as for the massless case with 0-bin subtractions \cite{Idilbi:2007ff}.

We compute the integrals $I_n, I_{\bar{n}}$ and $I_s$ by first carrying out the
$k^+$-integration with the method of residues and then the $k_\perp$-integration. So
the integrals $I_n$ and $I_{\bar{n}}$ are not treated symmetrically and lead to
different analytic expressions for the remaining $k^-$~integral. In the
$k^-$~integration all 
singularities cancel properly in the sum of all terms and the result is finite
in dimensional regularization. 
We start from the integral $I_n$ given in Eq.~(\ref{eq:defin}). Carrying out the
$k^+$- and $k_\perp$-integration as described above we obtain
\be
I_n= \frac{i}{(4\pi)^{d/2}}\,\Gamma\left(2-\frac{d}{2}\right)(M^2)^{d/2-2}
\,\int_0^1{dz\;\frac{(1-z)^{d/2-1}}{z}}
\label{eq:infinal}
\ee
with $z\equiv k^-/Q$. The integral in Eq.~(\ref{eq:infinal}) is divergent showing
that dimensional regularization fails. For the $I_{\bar{n}}$ integral it is
convenient to split it up as
\begin{equation}
I_{\bar{n}}=\tilde{I}_{\bar{n}}-\tilde{\mu}^{-2\epsilon}B_0(0,M^2,0)\,,
\label{eq:inbarsplitting}
\end{equation}
where $B_0(p^2,m^2_1,m^2_2)$ denotes the usual two-point function. After
performing the $k^+$- and $k_\perp$-integrations we arrive at ($x^2= M^2/(Q^2+i0)$)
\ba
\tilde{I}_{\bar{n}}&=&\frac{i}{(4\pi)^{d/2}}\,
\Gamma\left(2-\frac{d}{2}\right)(M^2)^{d/2-2}
\nn\\
&&\times\int_0^{\infty}{dz\frac{1-\left(\frac{z}{-x^2}\right)^{d/2-2}}{x^2+z}}
\,.
\label{eq:inbarfinal}
\ea
The result differs from Eq.~(\ref{eq:infinal}) since the
integrations for $I_n$ and $I_{\bar{n}}$ were performed in an asymmetric way. 
Finally we compute $I_s$ given in Eq.~(\ref{eq:defis}) in the same way and
obtain 
\begin{equation}
I_s=\frac{i}{(4\pi)^{d/2}}
\,\Gamma\left(2-\frac{d}{2}\right)(M^2)^{d/2-2}
\,\int_0^\infty{\frac{dz}{z}}\,,
\label{eq:isfinal}
\end{equation}
which is again not regularized by dimensional regularization. 

Upon summing all contributions according to Eq.~(\ref{eq:Isum}) the singularities
cancel exactly. For this purpose we split $I_s$ given in Eq.~(\ref{eq:isfinal})
into an integration from $0$ to $1$ and from $1$ to   
$\infty$. The first integral is then recombined with $I_n$, the second with
$\tilde{I}_{\bar{n}}$. For the sum of the divergent integrals we then obtain
\begin{align}
& \int_0^1{\frac{(1-z)^{d/2-1}-1}{z}dz}
+\int_0^1{\frac{1-\left(\frac{z}{-x^2}\right)^{d/2-2}}{x^2+z}dz}
 \nn\\
&-\int_1^\infty{\frac{x^2+z\left(\frac{z}{-x^2}\right)^{d/2-2}}{(x^2+z)z}dz}=\nn\\
&-H_{\frac{d}{2}-1}
+\Gamma\left(\frac{d}{2}\right)\Gamma\left(1-\frac{d}{2}\right)
-\ln{\left(-x^2\right)} \, ,
\label{eq:ininbaris}
\end{align}
where $H_{a}$ denotes the Harmonic number function.

Finally, we compute the wave function renormalization diagrams $W_{f,a}$ and
$W_{f,b}$ for the external massless collinear quarks. The sum of both
self-energy diagrams for arbitrary collinear external momentum $p^\mu$ can be
readily combined and reads 
\begin{align}
% W_{f,a} & = - g^2 C_F (d-2) i \frac{\slash{\bar{n}}}{2}\tilde{\mu}^{2\epsilon}
% \int{\frac{d^dk}{(2\pi)^d}}\frac{-k_\perp^2}{[p^-+k^-][(p+k)^2+i
% \epsilon][k^2-M^2+i \epsilon]} \, , \\ 
% W_{f,b} & = - g^2 C_F (d-2) i \frac{\slash{\bar{n}}}{2}\tilde{\mu}^{2\epsilon}
% \int{\frac{d^dk}{(2\pi)^d}\frac{1}{[p^-+k^-][k^2-M^2+i \epsilon]}} \, . \\
 W_{f,a} + & W_{f,b}  =
 - g^2 C_F (d-2) i \frac{\slash{\bar{n}}}{2}\tilde{\mu}^{2\epsilon} \frac{1}{Q}
 \int{\frac{d^dk}{(2\pi)^d}}\nn\\
&\frac{p^2+Qk^+}{[p^2(1+k^-/Q)+Qk^{+}+
   k^2+i\se][k^2-M^2+i \epsilon]}\,,
\label{eq:Wabfull}
\end{align}
which is known to reproduce full theory self energy graph.
The wave-function renormalization contribution can be identified in the limit
$p^2\to 0$ (and $\epsilon\to 0$) giving
\begin{align}
 W_{f,a} +  W_{f,b} & \stackrel{p^2\to 0}{\longrightarrow}
 i \frac{\slash{\bar{n}}}{2} \frac{p^2}{p^-} W_f
\,,
\label{eq:Wabwavefct}
\end{align}
where
\begin{equation}
W_f = \frac{\alpha_s C_F}{4\pi}\left[\frac{1}{\epsilon}-\ln{\left(\frac{M^2}{\mu^2}\right)}-\frac{1}{2}\right] \, .
\label{eq:wavefunctionrenormalization}
\end{equation}
The soft mass mode bin subtraction for the wave function contribution turns out
to be power suppressed by $M/Q$. To demonstrate this let us first consider the
power counting of the collinear mass mode self-energy diagrams given in
Eq.~(\ref{eq:Wabfull}): the collinear loop momenta scale like
$k^\mu=(k^-,k^+,k_\perp)\sim (Q,M^2/Q,M)$ and the external massless collinear modes
interacting with the collinear mass modes thus obey the same scaling giving
$p^2\sim (p+k)^2\sim M^2$. This yields $W_{f,a}+ W_{f,b}\sim M^2/Q$ for the result in
Eq.~(\ref{eq:Wabfull}) and gives the same counting for the wave-function
contribution given in Eq.~(\ref{eq:Wabwavefct}), $p^2/Q\times W_f\sim M^2/Q$. This
demonstrates that $W_f\sim{\cal O}(1)$ as we have obtained also in the explicit result
shown in Eq.~(\ref{eq:wavefunctionrenormalization}). For the mass mode bin
subtraction to Eq.~(\ref{eq:Wabfull}) we have 
to apply the counting $k^\mu_{s}\sim (M,M,M)$ for the loop momentum and
this increases the typical invariant mass of the massless collinear quarks that
interact with the soft mass mode gauge bosons to $p^2\sim (p+k_s)^2\sim Q M$. The mass mode
bin integral that emerges therefore has the form of a simple tadpole graph,
\begin{align}
(W_{f,a} +  W_{f,b})_{0M} & =
 - g^2 C_F (d-2) i \frac{\slash{\bar{n}}}{2}\tilde{\mu}^{2\epsilon} \frac{1}{Q}
 \nn\\
&\times\int{\frac{d^dk}{(2\pi)^d}}\frac{1}{[k^2-M^2+i \epsilon]}\,.
\label{eq:Wabfullsoft}
\end{align}
with the counting
$(W_{f,a}+ W_{f,b})_{0M}\sim M^2/Q$. The same counting also applies to the
resulting wave-function renormalization contribution, i.e.\ 
$p^2/Q\times (W_f)_{0M}\sim M^2/Q$, and because $p^2\sim Q M$ we find that 
$(W_f)_{0M}\sim M/Q$. This shows that the mass mode bin subtraction for the
collinear wave function renormalization due to the mass modes belongs to a
subleading treatment beyond the scope of the treatment discussed here.\footnote{
Since linear $M/Q$-suppressed terms do not exist in the non-singular collinear terms
that can be obtained in the full theory calculation, the subleading effective
field theory treatment contains a mechanism that makes these contributions
vanish identically.
}
Combining Eqs.~(\ref{eq:inbarsplitting}), (\ref{eq:ininbaris}) and
(\ref{eq:wavefunctionrenormalization})  and expanding in $\epsilon$ we arrive
at the final result 
\begin{align}
\delta F_{\textrm{eff}}^{\rm bare}&(Q,M,\mu)=\nn\\
 =&2i g^2 C_F\left[\tilde{\mu}^{2\epsilon}\left(I_n+\tilde{I}_{\bar{n}}-I_s\right)-B_0(0,M^2,0)\right]-W_f\nn\\
=&\frac{\alpha_s C_F}{4\pi}
\left\{\frac{2}{\epsilon^2}+\frac{3}{\epsilon}-\frac{2}{\epsilon}\,\ln\left(\frac{-Q^2}{\mu^2}\right)+\ln\left(\frac{M^2}{\mu^2}\right)
\right.\nn\\
&\times\left.\left[2 \, \ln\left(\frac{-Q^2}{\mu^2}\right)-\ln\left(\frac{M^2}{\mu^2}\right)-3\right] -\frac{5\pi^2}{6}+\frac{9}{2} \right\}\,.
\label{eq:effectivesumddim}
\end{align}
In Eq.~(\ref{eq:effectivesumddim}) the UV-divergences do not depend on the mass
and agree exactly with those of the sum of the corresponding collinear and soft
effective theory diagrams for massless gluons, see Eq.~(\ref{eq:Zc}). After
renormalization we thus obtain the result for $\delta F_{\textrm{eff}}$ given in
Eq.~(\ref{eq:hardeffm}). 
For $\mu_H\sim Q$ the large logarithms that occur in $\delta
F_{\rm eff}$ for $M \ll Q$ cancel entirely with the corresponding logarithms in the full theory form factor $\delta F_m$ in Eq.~(\ref{eq:hardQCDmexpanded}).

\subsubsection{Current mass mode matching coefficient}

The virtual corrections contained in $\delta F_{\rm eff}$ coincide with the
mass-shell contributions that are contained in the mass mode matching
coefficient ${\cal M}_H$. The latter arises when the mass modes are integrated
out in the hard current RG-evolution at the scale
$\mu_M\sim M$.  When $\mu_M$ is not exactly equal to $M$,  ${\cal M}_H$
contains a large "rapidity" logarithm that is known to
exponentiate~\cite{Chiu:2007dg,Becher:2010tm}. The summation of these logarithms can be carried
out independently after the $\mu$-evolution has been settled, and this is
the approach we are adopting here. In Ref.~\cite{Chiu:2011qc,Chiu:2012ir}
it has been pointed out that $\mu$-evolution and the summation of
"rapidity" logarithms can also be carried out simultaneously via a
two-dimensional evolution merging renormalization evolution in virtuality
(within dimensional regularization) with renormalization evolution in
rapidity. In the following we outline the corresponding calculations in
anticipation of analogous computations for the jet and the soft functions.
We employ an analytic "$\alpha$-regulator" for the $k^-$ integrations
\be\label{eq:regulator}
\frac{dk^-}{k^-}\,\to\, dk^-\frac{\nu^\alpha}{(k^-)^{1+\alpha}} \,.
\ee
which is equivalent to setting $dz/z \to dz/z^{1+\alpha}(\nu/Q)^\alpha$ in Eqs.~(\ref{eq:infinal}),~(\ref{eq:inbarfinal}) and (\ref{eq:isfinal}). In this
context the scale $\nu$ is an auxiliary scale to maintain the dimensions
of the regulated integrals which adopts a similar role as the $\mu$ scale
in dimensional regularization. In particular, also the strong coupling
adopts a $\nu$ scaling proportional to $\alpha$.
The $\alpha$-regulator was discussed in Ref.~\cite{Becher:2011dz} and treats the $n$-
and $\bar n$-collinear sectors as well as the soft boundaries towards the
two collinear sectors in an asymmetric way. Upon taking the limits $\alpha\to 0$ prior to
$\epsilon\to 0$, we derive a $\nu$-evolution equation for the current from
renormalizing  the $1/\epsilon$ and $1/\alpha$ singularities. Note that the
$\alpha$-regulator as defined in Eq.~(\ref{eq:regulator}) entails the
occourence of spurious $\ln{(Q)}$-terms that are not supposed to be confused
with the analytic $\ln{(-Q^2)}$-terms. Only the latter are altered when
continuing to the Euclidean region. 

We write 
\begin{align}
{\cal M}_H= 1+{\cal M}_{H,n}^{(1)}+{\cal M}_{H,\bar n}^{(1)}+{\cal M}_{H,s}^{(1)}
\end{align}
for the one-loop $n$- and $\bar n$-collinear and soft contributions. 
With the $\alpha$-regulator the soft mass mode as well as the soft-bin
mass mode subtraction diagrams lead to vanishing scaleless integrals, i.e ${\cal M}_{H,s}^{(1)}=0$.\footnote{We note that the result for the soft mass mode and the soft-bin
contributions depends on the regulator and does not vanish in general.} We
obtain for the $n$-collinear contribution ($\alpha_s=\alpha_s(\mu_M)$)
\begin{align}
&\mathcal{M}_{H,n}^{(1)} =  \,\frac{\alpha_s
C_F}{4\pi}\left\{\frac{2}{\alpha \epsilon}-\frac{2}{\alpha}\,\ln \left(\frac{M^2}{\mu_M^2}\right)+\frac{2}{\epsilon}\,\ln\left(\frac{\nu}{Q}\right)+\frac{3}{2\epsilon}\right.\nn \\
&\left.-2\,\ln\left(\frac{M^2}{\mu_M^2}\right)\ln\left(\frac{\nu}{Q}\right)
  -\frac{3}{2}\,\ln\left(\frac{M^2}{\mu_M^2}\right)+\frac{9}{4}-\frac{\pi^2}{3} \right\},\label{eq:M_Hn}
\end{align}
and for the $\bar{n}$-collinear contribution ($Q^2\equiv Q^2+i 0$)
\begin{align}
&\mathcal{M}_{H,\bar{n}}^{(1)}= \, \frac{\alpha_s
C_F}{4\pi}\left\{-\frac{2}{\alpha\epsilon}+\frac{2}{\alpha}\,\ln\left(\frac{M^2}{\mu_M^2}\right)+\frac{2}{\epsilon^2}\right.\nn \\
& \hspace{0.2cm}-\frac{2}{\epsilon}\left[\ln\left(\frac{-Q^2}{M^2}\right)+\ln\left(\frac{\nu}{Q}\right)\right]-\frac{2}{\epsilon}\,\ln\left(\frac{M^2}{\mu_M^2}\right)+ \frac{3}{2\epsilon}\nn\\
& \hspace{0.2cm} +\left.\ln^2
\left(\frac{M^2}{\mu_M^2}\right)+2\,\ln\left(\frac{M^2}{\mu_M^2}\right)\left[\ln\left(\frac{-Q^2}{M^2}\right)+\ln\left(\frac{\nu}{Q}\right)\right]\right.\nn \\
&\hspace{0.2cm} -\left.\frac{3}{2}\,\ln\left(\frac{M^2}{\mu_M^2}\right)+\frac{9}{4}-\frac{\pi^2}{2}\right\}\,. \label{eq:M_Hnbar}
\end{align}
We see that  ${\cal M}_{H,n}^{(1)}$ is free of large logarithms for $\nu=\nu_n\sim
\mu_H\sim Q$, and ${\cal M}_{H,\bar n}^{(1)}$ is free of large logarithms for
$\nu=\nu_{\bar n}\sim \mu_M^2/\mu_H\sim M^2/Q$. It is then possible to set up
an evolution in $\nu$ between $\nu_{\bar n}$ and $\nu_n$ interpreting
${\cal M}_{H,\bar n}$ as a  "low-scale" effective theory contribution that is
being renormalized. In this context the finite terms of ${\cal M}_{H,n}$
represent a "large-scale" matching contribution. We emphasize that we use
this interpretation merely as a practical guide, because the corresponding
physical implications are subtle and strongly regulator-dependent.  The
resulting current renormalization constant reads
\begin{align}
&Z_{\bar n} (Q,M,\mu_{M},\nu)=\,1+\frac{\alpha_s C_F}{4\pi}\left\{-\frac{2}{\alpha \epsilon}+\frac{2}{\alpha}\,\ln\left(\frac{M^2}{\mu_{M}^2}\right)\right.\nn\\
&\hspace{0.5cm}+\left.\frac{2}{\epsilon^2}-\frac{2}{\epsilon}\left[\ln\left(\frac{-Q^2}{M^2}\right)+\ln\left(\frac{\nu}{Q}\right)\right] \right.\nn \\
&\hspace{0.5cm}-\left.\frac{2}{\epsilon}\,\ln\left(\frac{M^2}{\mu_M^2}\right)+\frac{3}{2\epsilon}\right\}\, .
\end{align}
With $d\alpha_s/d\,\ln\,\nu=-\alpha\,\alpha_s$ and ${\cal M}_{H,\bar n}=1+{\cal M}_{H,\bar n}^{(1)}$ we obtain the $\nu$-evolution equation
\begin{align}
&\frac{d}{d \,\ln\,\nu} {\cal M}_{H,\bar n}(Q,M,\mu_M,\nu)  \nn \\
&=\bigg[ \frac{\alpha_s}{4\pi} \, \frac{\Gamma_0}{2} \, \ln\bigg( \frac{M^2}{\mu_M^2}  \bigg) \bigg] {\cal M}_{H,\bar n} (Q,M,\mu_M,\nu)\,, \label{eq:nuevolutioncurrent}
\end{align} 
where the dependence on the cusp anomalous dimension on the RHS is related
to the path-independence of the evolution in
$\mu$-$\nu$-space, which holds to all orders~\cite{Chiu:2012ir}. It can be also understood from the fact
that for $\mu_M=M$ the corresponding logarithms are being summed by the $\mu$-evolution
factors and there is no $\nu$-evolution.
Solving Eq.~(\ref{eq:nuevolutioncurrent}) and expanding out the terms that do not
involve large logarithms leads to 
\begin{align}
&\mathcal{M}_H(Q,M,\mu_M,\nu_n,\nu_{\bar n}) = 
\exp\bigg[ \frac{\alpha_s} {4\pi}\, \frac{\Gamma_0}{2}\,\ln\bigg( \frac{M^2}{\mu_M^2} \bigg) \,\ln\bigg( \frac{\nu_n}{\nu_{\bar n}}  \bigg) \bigg]
\nn\\
&\times\left(1+\frac{\alpha_sC_F}{4\pi}\left\{-\ln{\left(\frac{M^2}{\mu^2_M}\right)}\left[\ln{\left(\frac{\nu^2_n}{-Q^2}\right)}-\ln{\left(\frac{-Q^2 \nu^2_{\bar n}}{M^4}\right)}\right]\right.\right.\nn\\
&\hspace{0.5cm}\left.+\left.\ln^2{\left(\frac{M^2}{\mu^2_M}\right)}-3\, \ln{\left(\frac{M^2}{\mu^2_M}\right)}+\frac{9}{2}-\frac{5}{6}\pi^2 \right\}\right) \, .
\end{align}
It is convenient to adopt the choices $\nu_n=\mu_H$ and $\nu_{\bar
  n}=\mu_M^2/\mu_H$, which then gives Eq.~(\ref{eq:matchingII_RRG}). We note
that $\nu_{\bar{n}}$ might also be chosen complex to sum additional $i\pi$ terms.

\subsection{Jet Function and Collinear Mass Mode Matching Coefficient}\label{sect:jetfunction}

\subsubsection{Jet function}

\begin{figure}
 \centering
 \includegraphics[width=\linewidth]{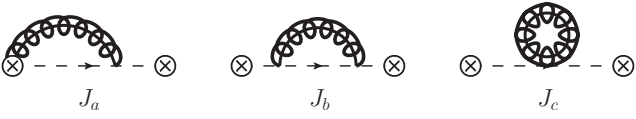}
 \caption{Non-vanishing EFT diagrams for the computation of the jet function. The required
   soft mass mode bin subtractions are implicit. Concerning $J_a$ also the
   right-symmetric diagram has to be taken into
   account.\label{fig:jetfunction_diagrams}}  
\end{figure}

In this section we calculate the ${\cal O}(\alpha_s)$ mass mode gauge boson
contribution to the 
$n$-collinear massless quark jet function $J_{n,m}(s,M,\mu)$ and relate it
to the thrust jet function of
Eq.~(\ref{eq:jetmassive}) appearing in the factorization 
theorems~(\ref{eq:diffsigmaIII}) and (\ref{eq:diffsigmaIV}) for scenarios~III
and IV where $\lambda>\lambda_M$. The diagrams contributing to the $n$-collinear
jet function are listed in Fig.~\ref{fig:jetfunction_diagrams}. Similarly to
the hard function calculation, the singularities that arise in the collinear
mass mode contributions and which are not handled by dimensional regularization
are canceled by the soft-bin mass mode subtractions.

The mass mode contributions to the jet function arise in the situation
$\lambda\geq \lambda_M$, where the collinear mass modes can yield real
  and virtual effects with typical invariant mass $Q\lambda$. So we can use the
scaling $k^\mu\sim Q(\lambda^2,1,\lambda)$ for the $n$-collinear mass mode gauge
bosons with the assignment $\lambda_M\sim\lambda$ to keep the full mass dependence\footnote{ In scenario~III the soft mass-shell fluctuations have typical momenta
$k\sim Q(\lambda_M,\lambda_M,\lambda_M)$. In scenario~IV, where $M$ is
      below the soft scale, they become irrelevant for the separation
      of modes, so the soft mass
modes only have typical momenta $k\sim Q(\lambda^2,\lambda^2,\lambda^2)$.}. The jet momentum and its invariant mass are denoted with $p^\mu$ and $s\equiv
p^2=p^-p^+=Qp^+$, the prescription $s\rightarrow s+i0$ is understood. We keep $p^+$ explicitely non-zero in the subsequent
computations to account for the non-vanishing off-shellness $s$ of the jet
sector and use a frame with $p_\perp=0$.

Taking into account the traces and the prefactor of the jet function matrix element in
Eq.~(\ref{eq:jetfunctiondefinition}) we obtain for diagram $J_a$   
\begin{align}
J_a=&-\frac{2g^2 C_F}{\pi s} \tilde{\mu}^{2\epsilon}
\int{\frac{d^dk}{(2\pi)^d}\frac{p^--k^-}{[(p-k)^2+i\se]}}
\nn\\&\hspace{1cm}\times
\frac{1}{[k^-+i\se][k^2-M^2+i\se]}  \, .
\label{eq:Ja}
\end{align}
For our choice of the external momenta this yields 
\begin{align}
J_a =&-\frac{2g^2 C_F}{\pi s} \tilde{\mu}^{2\epsilon}
\int{\frac{d^dk}{(2\pi)^d}\frac{Q-k^-}{[s(1-k^-/Q)-Qk^+ +k^2+i\se]}}
\nn\\
&\hspace{1cm}\times\frac{1}{[k^-+i\se][k^2-M^2+i\se]} \nn \\[2mm]
& \equiv -\frac{2g^2 C_F}{\pi s} \tilde{\mu}^{2\epsilon}\, I_a \, .
\label{eq:Ia}
\end{align}
The corresponding soft mass mode bin subtraction with the scaling
$k^\mu\sim Q(\lambda_M,\lambda_M,\lambda_M)$ for scenario~III (or
$k^\mu\sim Q(\lambda^2,\lambda^2,\lambda^2)$ for scenario~IV) 
gives
\begin{align}
J_{a,0M} &=-\frac{2g^2 C_F}{\pi s} \tilde{\mu}^{2\epsilon}
\int{\frac{d^dk}{(2\pi)^d}\frac{Q}{[s-Qk^+ +i\se][k^-+i\se]}}\nn\\&
\hspace{1cm}\times\frac{1}{[k^2-M^2+i\se]} \nn \\[2mm]
& \equiv -\frac{2g^2 C_F}{\pi s} \tilde{\mu}^{2\epsilon}\, I_{a,0M} \, .
\label{eq:Ia0M}
\end{align}
Note that for scenario~III the invariant mass term $s$ is included as an
infrared scale to account for the non-zero jet-invariant mass. This is
  related to the fact that the invariant mass of the massless collinear quark
  increases from $Q\lambda$ to $Q \sqrt{\lambda_{M}}$ by an interaction
    with a virtual soft mass mode which also changes the parametric counting
  of $s$. In close analogy to the wave function contributions to the vertex corrections 
discussed in Sec.~\ref{sect:hardfunction} the sum of the self-energy diagrams
$J_b$ and $J_c$ reads 
\begin{align}
J_b + J_c =&-\frac{g^2 C_F (d-2)}{\pi s} \tilde{\mu}^{2\epsilon}
\int{\frac{d^dk}{(2\pi)^d}}\frac{1}{[k^2-M^2+i\se]}\nn\\
&\hspace{1cm}\times\frac{1+Qk^+/s}{[s(1+k^-/Q)+Qk^+
    +k^2+i\se]} \nn\\[2mm] 
 \equiv& -\frac{g^2 C_F (d-2)}{\pi s} \tilde{\mu}^{2\epsilon} \, (I_b +I_c) \,.
\label{eq:Ibc}
\end{align}
As we have discussed for the vertex corrections in Sec.~\ref{sect:hardfunction},
the soft mass mode bin subtraction terms to the self-energy diagrams belong
to a subleading effective theory treatment and are thus not considered.

To compute the integrals, we apply the same technique as in
Sec.~\ref{sect:hardfunction}. So we first carry out the $k^+$ and $k_\perp$
integrations and combine the contributions of all diagrams prior to the  $k^-$
integration. For $I_a$ we then arrive at 
\begin{align}
I_a=&\frac{i}{(4\pi)^{d/2}}\Gamma\left(2-\frac{d}{2}\right)(M^2)^{d/2-2}
\nn\\
&\times\int_0^1{\frac{dz}{z}(1-z)^{d/2-1}\left(1-\frac{s}{M^2}z\right)^{d/2-2}}
\label{eq:divergentindreg}
\end{align}
with $z\equiv k^-/Q$. There is an unregularized divergence for $z\to 0$,
which will be cured by the corresponding soft mass mode bin subtraction, 
\begin{align}
I_{a,0M}=\frac{i}{(4\pi)^{d/2}}&\Gamma\left(2-\frac{d}{2}\right)(M^2)^{d/2-2}\nn\\
&\times\int_0^\infty{\frac{dz}{z}\left(1-\frac{s}{M^2}z\right)^{d/2-2}} \, .
\label{eq:softmassmodesubtractioninjc}
\end{align}
Subtracting Eq.~(\ref{eq:softmassmodesubtractioninjc}) from
Eq.~(\ref{eq:divergentindreg}) yields 
\begin{align}
I_a-I_{a,0M} \sim
&\left\{\int_0^1{\frac{dz}{z}\left[(1-z)^{d/2-1}-1\right]\left(1-\frac{s}{M^2}z\right)^{d/2-2}}\right.
\nn\\
&\left.-\int_1^\infty{\frac{dz}{z}\left(1-\frac{s}{M^2}z\right)^{d/2-2}}\right\} \, ,
\label{eq:jcsubtracted}
\end{align}
which is finite in dimensional regularization. We note that the rapidity divergences we encounter only affect the virtual (distributive) contribution to the jet function. As made explicit in Eq.~(\ref{eq:MJ_0M}) the soft mass mode bin contributions do not lead to any real radiative terms, consistently with the counting argument given after Eq.~(\ref{eq:Ia0M}).

Next, we proceed to the calculation of the self energy integrals
$I_b+I_c$. Defining $z\equiv -k^+/p^+$ we arrive at the expression 
\begin{align}
I_{b}+I_{c}=&\frac{i}{(4\pi)^{d/2}}\Gamma\left(2-\frac{d}{2}\right)
(M^2)^{d/2-2}\nn\\&\times
\int_0^1{dz\;(1-z)^{d/2-1}\left(1-\frac{s}{M^2}z\right)^{d/2-2}} \, .
\label{eq:resultijeplusf}
\end{align}

The sum of all diagrams in the limit $\epsilon\rightarrow0$ yields
\begin{align}
&2 J_a + J_b + J_c - 2J_{a,0M}= \nn \\
& = \frac{\alpha_sC_F}{4\pi}\frac{i\pi}{s} \left\{\frac{4}{\epsilon^2} +\frac{1}{\epsilon}\left[3-4\,\ln{\left(\frac{M^2}{\mu^2}\right)}-4\,\ln{\left(\frac{-s}{M^2}\right)}\right] \right.\nn \\
& \hspace{0.5cm} -4 \,{\rm Li}_2\left(\frac{s}{M^2}\right)+ 2\,\ln^2{\left(\frac{-s}{\mu^2}\right)}-2\,\ln^2{\left(\frac{-s}{M^2}\right)} \nn\\
& \hspace{0.5cm} -3\,\ln{\left(\frac{M^2}{\mu^2}\right)}+\frac{(M^2-s)(3s+M^2)}{s^2}\,\ln{\left(1-\frac{s}{M^2}\right)} \nn \\
& \hspace{0.5cm}+\left.\frac{M^2}{s} +7-\pi^2\right\} \,.
\end{align}
We can take the absorptive part with the help of the relations in the
appendix of Ref.~\cite{Fleming:2007xt}, which leads to the unrenormalized mass
mode contributions to the jet function
\begin{align}
& \mu^2 \delta J_{n,m}^{\rm bare}(s,M,\mu) = \frac{\alpha_sC_F}{4\pi}\left\{\delta(\bar{s})\left(\frac{4}{\epsilon^2}+\frac{3}{\epsilon}\right) -\frac{4}{\epsilon}\left[\frac{\theta(\bar{s})}{\bar{s}}\right]_+ \nn \right.\\
& \hspace{0.5cm}+\delta(\bar{s})\left[-2\,\ln^2\left(\frac{M^2}{\mu^2}\right)-3\,\ln\left(\frac{M^2}{\mu^2}\right)+\frac{9}{2}-\pi^2\right]\nn\\
& \hspace{0.5cm}+ 4 \,\ln\left(\frac{M^2}{\mu^2}\right)\left[\frac{\theta(\bar{s})}{\bar{s}}\right]_{+} +\mu^2\theta\left(s-M^2\right) \nn \\
& \hspace{1.0cm} \left. \times \left[\frac{(M^2-s)(3s+M^2)}{s^3}+\frac{4}{s}\,\ln{\left(\frac{s}{M^2}\right)}\right]\right\}
\, .\label{eq:Jbare}
\end{align}
We see that the UV-divergences are mass-independent and agree with those from
the purely massless jet function.  
Multiplying the result by a factor of $2$ to account for the combination of the
two hemisphere jet function into the thrust jet function and using the jet
function renormalization counterterm given in Eq.~(\ref{eq:ZJ}) we get $\delta
J_{m}(s,M,\mu)$ in Eq.~(\ref{eq:jetmassive}).  For $\mu^2=\mu_J^2 \sim s\sim Q^2\lambda^2$ all large logarithms that
arise for $M^2 \ll s$ cancel between the real and virtual mass mode
contributions.

\subsubsection{Collinear mass mode matching coefficient}

The contributions to the collinear mass mode matching coefficient ${\cal M}_J$ arise from the virtual (distributive) corrections, see Eq.~(\ref{eq:MJet}). 
In the following we discuss the calculations to sum the large logarithms
that arise
for $\mu_M\sim M$, when $\mu_M$ is not equal to $M$, see Eq.~(\ref{eq:MJet_0}). We
closely follow the method outlined in Sec.~\ref{sect:hardfunction}. We write the convolution ${\cal M}_J = {\cal M}_{J,n} \otimes {\cal M}_{J,\bar{n}}$ as ($\tilde{s}=s/\mu_J^2$)
\begin{align}
 {\cal M}_{J} &= \delta(\tilde{s})+2{\cal M}_{J,M}^{(1)}+2{\cal M}_{J,0M}^{(1)}
\end{align}
for the one-loop virtual collinear and soft-bin mass mode contributions. Note the opposite sign convention for the latter compared to the jet function calculation.

Using the $\alpha$-regulator the collinear mass mode virtual contributions
read ($\alpha_s=\alpha_s(\mu_M)$)

\begin{align}
 &\mu_{J}^2 {\cal M}_{J,M}^{(1)} = \, \frac{\alpha_s
C_F}{4\pi}\left\{\delta(\tilde{s})\left[-\frac{4}{\alpha \epsilon}+\frac{4}{\alpha}\,\ln\left(\frac{M^2}{\mu_M^2}\right)\right.\right.\nn\\
& \hspace{0.5cm}-\left.\left.\frac{4}{\epsilon}\,\ln\left(\frac{\nu}{Q}\right)-\frac{3}{\epsilon}+4\,\ln\left(\frac{M^2}{\mu_M^2}\right)\ln\left(\frac{\nu}{Q}\right)
\right. \right. \nn \\
&\hspace{0.5cm}+\left.\left.3\, \ln\left(\frac{M^2}{\mu_M^2}\right)-\frac{9}{2}+\frac{2\pi^2}{3} \right] \right\}, 
\end{align}
and the contributions due to the soft-bin subtractions yield
\begin{align}
& \mu_{J}^2 {\cal M}_{J,0M}^{(1)}= \,- \frac{\alpha_s
C_F}{4\pi}\left\{\delta(\tilde{s})\left[-\frac{4}{\alpha
\epsilon}+\frac{4}{\alpha}\,\ln\left(\frac{M^2}{\mu_M^2}\right)+\frac{4}{\epsilon^2}\right.\right.
\nn\\
& \hspace{0.5cm}-\frac{4}{\epsilon}\,\ln\left(\frac{\nu \mu_{J}^2}{Q M^2}\right)-\frac{4}{\epsilon}\,\ln\left(\frac{M^2}{\mu_M^2}\right)+2\,\ln^2 \left(\frac{M^2}{\mu_M^2}\right)\nn\\
& \hspace{0.5cm}+\left.4\,\ln\left(\frac{M^2}{\mu_M^2}\right)\,\ln\left(\frac{\nu\mu_{J}^2}{Q M^2}\right)-\frac{\pi^2}{3}\right]\nn\\
 & \hspace{0.5cm}+\left.\left[\frac{\theta(\tilde{s})}{\tilde{s}}\right]_{+}\left[-\frac{4}{\epsilon}+4\,\ln\left(\frac{M^2}{\mu_M^2}\right)\right]
\right\} \, . \label{eq:MJ_0M}
\end{align}
We see that  ${\cal M}_{J,M}^{(1)}$ is free of large logarithms for $\nu=\nu_n\sim
\mu_H\sim Q$, and ${\cal M}_{J,0M}^{(1)}$ is free of large logarithms for
$\nu=\nu_{n,0}\sim \mu_H\mu_M^2/\mu_J^2\sim Q M^2/s$.   The renormalization constant for the "low-scale"
  contribution to the collinear mass mode matching coefficient  ${\cal
    M}_{J,0M}=\delta(\tilde{s})+2 {\cal M}_{J,0M}^{(1)}$ reads 
\begin{align}
&\mu_{J}^2 Z_{J,0M}(s,M,Q,\mu_{M},\nu)=\delta(\tilde{s})+\frac{\alpha_s C_F}{4\pi}\left\{\delta(\tilde{s})\left[\frac{8}{\alpha\epsilon} \right. \right.\nn \\
&\hspace{0.2cm}\left.\left.-\frac{8}{\alpha}\,\ln\left(\frac{M^2}{\mu_M^2}\right)-\frac{8}{\epsilon^2}+\frac{8}{\epsilon}\,\ln{\left(\frac{\nu\mu_{J}^{2}}{Q\mu_M^{2}}\right)}\right]+\left[\frac{\theta(\tilde{s})}{\tilde{s}}\right]_+ \frac{8}{\epsilon}\right\} \, ,
\end{align}
from which we obtain the $\nu$-evolution equation
\begin{align}
\label{eq:nuevolutionjet}
&\frac{d}{d\,\ln\,\nu} {\cal M}_{J,0M}(s,M,Q,\mu_M,\nu) \nn \\
&=\bigg[- \frac{\alpha_s}{4\pi} \,2\, \Gamma_0 \, 
\ln\bigg( \frac{M^2}{\mu_M^2}  \bigg) \bigg] {\cal M}_{J,0M} (s,M,Q,\mu_M,\nu)
\,.
\end{align} 
As for the current mass mode matching coefficient, there is no
$\nu$-evolution for $\mu_M=M$. Solving Eq.~(\ref{eq:nuevolutionjet}), which leads to the anticipated
exponentiation, and expanding out the terms that do not involve large
logarithms leads to
\begin{align}
&\mu_J^2 {\cal M}_{J}(s,M,\mu_M,\nu_n,\nu_{n,0}) \nn\\
&=\exp\bigg[-\frac{\alpha_s} {4\pi} \,2\,\Gamma_0\,\ln\bigg( \frac{M^2}{\mu_M^2}
\bigg) \ln\bigg( \frac{\nu_n}{\nu_{n,0}}  \bigg) \bigg] \left(1+\frac{\alpha_s C_F}{4\pi}\right.\nn \\
&\hspace{0.5cm}\times\bigg\{ \delta{(\tilde{s})} \left[8\,\ln{\left(\frac{M^2}{\mu^2_M}\right)}\left(\ln{\left(\frac{\nu_{n}}{\mu_H}\right)}-\ln{\left(\frac{\nu_{n,0} \mu^2_J}{\mu_H \mu_{M}^{2}}\right)}\right)\right.\nn\\
& \hspace{0.5cm}\left.\left.+\left.4\,\ln^2{\left(\frac{M^2}{\mu^2_M}\right)}+6\,\ln{\left(\frac{M^2}{\mu^2_M}\right)}-9+2\pi^2\right]\right.\right. \nn \\
& \hspace{0.5cm}+\left.\left.\left[\frac{\theta{(\tilde{s})}}{\tilde{s}}\right]_+\left[-8\,\ln{\left(\frac{M^2}{\mu_M^2}\right)}\right]\right\}\right)
\,
\end{align}
for the $n$-collinear matching coefficient.
It is convenient to adopt the choices $\nu_n=\mu_H$ and
$\nu_{n,0}=\mu_H\mu_M^2/ \mu_J^2$, which then gives  Eq.~(\ref{eq:MJet_RRG}).

\subsection{Mass Mode Contributions to the Soft Function}\label{sect:softfunction}

\subsubsection{Soft function}
\begin{figure}
 \centering
 \includegraphics[width=0.8\linewidth]{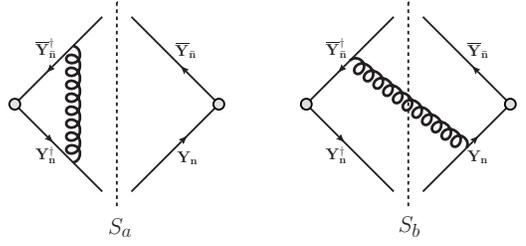}
 \caption{Non-vanishing Feynman diagrams for the computation of the one-loop mass mode
   contributions to the thrust soft function. The corresponding symmetric
   configurations are implied.\label{softfunction_diagrams}} 
\end{figure}

In this section we calculate the ${\cal O}(\alpha_s)$ contributions of the soft mass mode gauge
bosons to the soft function for scenario~IV, where $\lambda^2>\lambda_M$ and the
mass mode scale is below the ultrasoft scale. The non-vanishing diagrams are displayed in
Fig.~\ref{softfunction_diagrams}. Diagram $S_a$ ($S_b$) corresponds to virtual (real)
corrections. Their contributions to the thrust
soft function read 
\begin{align}\label{soft_start}
 S_a = & \, -2 i g^2 C_F \tilde{\mu}^{2\epsilon}  \int \frac{d^d k}{(2\pi)^d}
 \nn\\
& \hspace{1.3cm} \frac{\delta(\ell)}{(k^{+}+i\epsilon)(k^{-}-i\epsilon)(k^2-M^2+i\epsilon)} \, , \\
 S_b = & \, 4 \pi g^2 C_F \tilde{\mu}^{2\epsilon}\int \frac{d^d k}{(2\pi)^d}
 \, \Theta(k^{+}+k^{-})\delta(k^2-M^2)\nn\\
&\times \, \frac{\Theta(k^{-}-k^{+})\delta(\ell-k^{+})+\Theta(k^{+}-k^{-})\delta(\ell-k^{-})}{(k^{+}+i\epsilon)(k^{-}-i\epsilon)} \, .
\end{align}
Although the sum of all contributions is finite in dimensional regularization,
there are again individual singular regions with respect to large (or small) light cone momenta $k^+$ and $k^-$, or in other words with respect
to rapidity. These singularities cancel among themselves and no further recombination with other structures is required. We calculate the mass mode contributions to the soft function
with the $\alpha$-regulator already used in the Secs. ~\ref{sect:hardfunction}
and~\ref{sect:jetfunction}, since in this way the calculations become
  substantially more convenient.
The virtual diagrams $S_a$ eventually yield scaleless integrals and
give zero, similar to the case of a massless gluon exchange (see
Ref.~\cite{Fleming:2007xt}). The real radiation diagrams give two contributions,
\begin{equation}
 S_b= \delta S_{+} +\delta S_{-}
\end{equation}
related to the cases $k^+>k^-$ and $k^+<k^-$, which are treated asymmetrically with the $\alpha$-regulator. Applying the hemisphere constraints and performing the $k_\perp$-integration, one obtains the expressions 
\begin{align}
\,\delta S_{+}=  &\, \frac{\alpha_s C_F}{4\pi} \,4 \, \frac{\mu^{2\epsilon}
  e^{\gamma_E\epsilon}}{\Gamma(1-\epsilon)} \int_{\ell}^{\infty} dk^{+}\,
\theta(\ell\,k^{+}-M^2) \nn \\
& \times \frac{\nu^{\alpha}}{\ell^{\alpha}}\,\frac{\left(\ell\,k^{+}-M^2\right)^{-\epsilon}}{\ell\,k^{+}} \, , \\
\, \delta S_{-}= &\,\frac{\alpha_s C_F}{4\pi}\,4 \, \frac{\mu^{2\epsilon} e^{\gamma_E\epsilon}}{\Gamma(1-\epsilon)} \int_{\ell}^{\infty} dk^{-} \,\theta(\ell\,k^{-} -M^2) \nn \\
& \times \frac{\nu^{\alpha}}{\left(k^{-}\right)^{\alpha}}\,\frac{\left(\ell\,k^{-} -M^2\right)^{-\epsilon}}{\ell\,k^{-}} \, .
\end{align}
Performing the last integration and taking the limit $\epsilon\to 0$ after
$\alpha\to 0$ we obtain ($\bar{\ell}= \ell/\mu$)
\begin{align}
 \mu \,\delta S_{+}= \,& \frac{\alpha_s
C_F}{4\pi}\left\{\delta(\bar{\ell})\left[-\frac{4}{\alpha
\epsilon}+\frac{4}{\alpha}\,\ln\left(\frac{M^2}{\mu^2}\right) \right.\right.\nn \\
&\hspace{1cm}-\left.\frac{4}{\epsilon}\,\ln\left(\frac{\nu}{\mu}\right)+4\,\ln\left(\frac{M^2}{\mu^2}\right)\ln\left(\frac{\nu}{\mu}\right)\right]\nn\\
&\hspace{1cm}+\left[\frac{\theta(\bar{\ell})}{\bar{\ell}}\right]_{+}\left[\frac{4}{\epsilon}-4\,\ln\left(\frac{M^2}{\mu^2}\right)\right]\nn\\
&\hspace{1cm}-\left.\theta(\ell-M)\,\frac{4}{\bar{\ell}}\,\ln\left(\frac{\ell^2}{M^2}\right)\right\}
\, , \label{eq:deltaSplus}
\end{align}
and
\begin{align}
\mu \, \delta S_{-}= & \,\frac{\alpha_s C_F}{4\pi}\left\{\delta(\bar{\ell})\left[\frac{4}{\alpha\epsilon}-\frac{4}{\alpha}\,\ln\left(\frac{M^2}{\mu^2}\right)-\frac{4}{\epsilon^2} \right.\right.\nn\\
&\hspace{0.2cm}+\frac{4}{\epsilon}\,\ln\left(\frac{\nu \mu}{M^2}\right)+\frac{4}{\epsilon}\,\ln\left(\frac{M^2}{\mu^2}\right)-2\,\ln^2\left(\frac{M^2}{\mu^2}\right)\nn
\\
 &\hspace{0.2cm}-\left.4\,\ln\left(\frac{M^2}{\mu^2}\right)\ln\left(\frac{\nu \mu}{M^2}\right)+\frac{\pi^2}{3}\right]\nn\\
  &\hspace{0.2cm}+\left[\frac{\theta(\bar{\ell})}{\bar{\ell}}\right]_{+}\left[\frac{4}{\epsilon}-4\,\ln\left(\frac{M^2}{\mu^2}\right)\right]\nn\\
&\hspace{0.2cm}-\left.\theta(\ell-M)\,\frac{4}{\bar{\ell}}\,\ln\left(\frac{\ell^2}{M^2}\right)\right\} \, . \label{eq:deltaSminus}
\end{align}
In the sum all $\alpha$-singularities and the dependence on $\nu$ cancels,
and we obtain
\begin{align}
 \label{eq:softmassive4d}
 \mu \, \delta& S_{m}^{\rm bare}(\ell,M,\mu) =  \frac{\alpha_s C_F}{4\pi}\left\{-\frac{4}{\epsilon^2}\delta(\bar{\ell})
+\frac{8}{\epsilon}\left[\frac{\theta(\bar{\ell})}{\bar{\ell}}\right]_+
\right.
\nn\\&+\left.\delta(\bar{\ell}) \left[2 \,\ln^2\left(\frac{M^2}{\mu^2}\right) +\frac{\pi^2}{3}\right]-8\,\ln\left(\frac{M^2}{\mu^2}\right) \left[\frac{\theta(\bar{\ell})}{\bar{\ell}}\right]_{+}\right.\nn\\
&+ \left. \theta(\ell-M) \left[-\frac{8}{\bar{\ell}} \,\ln \left(\frac{\ell^2}{M^2} \right) \right]\right\}\,,
\end{align}
for the thrust soft function.
We see that the UV-divergences are mass-independent and agree exactly with the well
known divergences of the massless gluon contributions to the soft function.
Subtracting the divergences with the counterterm of Eq.~(\ref{eq:ZS}) leads to
the renormalized expression of the thrust soft function given in
Eq.~(\ref{eq:softmassive}).   
For $\mu=\mu_s \sim \ell \sim Q\lambda^2$ all large logarithms that arise
for $M \ll \ell$ cancel between the real and virtual mass mode
contributions. Finally we emphasize that for $M\to 0$ we obtain the results for the well-known
one-loop corrections for the massless soft function. This is achieved without
any subtraction as a manifastation that the mass-shell soft mass modes
do not have to be separated.

\subsubsection{Soft mass mode matching}

For the formulation of the factorization theorems we have adopted the
top-down convention, where the main renormalization scale $\mu$ is set equal
to the soft scale so that hard current and jet function both are evolved
to the soft scale.
It is also possible to set $\mu$ equal to the jet scale. In this situation
a different setup for the renormalization group evolution is realized,
where the hard current is evolved down and the soft function is evolved up
to the jet scale. For scenario~III there are then no mass mode matching
coefficients for the hard current and the jet function evolution. Rather,
the mass-shell mass mode contributions are "integrated in" at the scale
$\mu_M\sim M$ in the evolution of the soft function from the soft scale,
where the massless soft modes fluctuate, up to the jet scale. This is in
analogy to the generation of heavy quark parton distribution functions in
the ACOT-scheme. Consistency of the different choices for the main
renormalization scale $\mu$ then leads to the consistency
relation~(\ref{eq:smoothnessconstr}).

In this context it is also interesting to check the consistency
regarding the exponentiation of the large logarithms contained in mass
mode matching coefficients. In the just mentioned evolution scenario where
the soft function is evolved up to the jet scale, the soft mass mode
matching coefficient consists of the virtual corrections given in
Eqs.~(\ref{eq:softmassivevirt}) in the fixed-order expansion. This is also obvious from the fact that the soft
real radiation
contributions do not contribute in scenario~III.  In the following we
discuss the calculations to sum the large logarithms that arise
for $\mu_M\sim M$, when $\mu_M$ is not equal to $M$. We again closely
follow the method already used in Secs.~\ref{sect:hardfunction}
and~\ref{sect:jetfunction}.
The one-loop soft mass mode matching coefficient reads ($\tilde{\ell}=\ell/\mu_S$)
\begin{align}
 {\cal M}_S = \delta(\tilde{\ell})+ \delta S_+^{\rm virt} + \delta S_-^{\rm virt}
\end{align}
where $\delta S_+^{\rm virt}$ and $ \delta S_-^{\rm virt}$ denote the
  distributive terms in Eqs.~(\ref{eq:deltaSplus}) and~(\ref{eq:deltaSminus}). 
We see that  $\delta S_-^{\rm virt}$ is free of large logarithms for
  $\nu=\nu_-\sim \mu_M^2/\mu_S\sim M^2/\ell \sim M^2/Q\tau$, and $\delta
  S_+^{\rm virt}$ is free of large logarithms for $\nu=\nu_+\sim \mu_S\sim
  \ell\sim Q\tau$. The renormalization constant for the "low-scale"
  contribution to the soft mass mode matching coefficient ${\cal M}_{S,+}=\delta(\tilde{\ell})+ \delta S_+^{\rm virt}$ then reads 
\begin{align}
&\mu_S Z_+(\ell,M,\mu_M,\nu) =\,\delta(\tilde{\ell})+\frac{\alpha_s
C_F}{4\pi}\left\{\delta(\tilde{\ell})\left[-\frac{4}{\alpha \epsilon}
\right.\right.\nn\\
&\hspace{0.1cm}\left.\left.+\frac{4}{\alpha}\,\ln\left(\frac{M^2}{\mu_M^2}\right)-\frac{4}{\epsilon}\,\ln\left(\frac{\nu}{\mu_S}\right)\right]+\left[\frac{\theta(\tilde{\ell})}{\tilde{\ell}}\right]_{+}\frac{4}{\epsilon}\right\} \, .
\end{align}
We obtain the $\nu$-evolution equation ($\alpha_s=\alpha_s(\mu_M)$)
\begin{align}
\label{eq:nuevolutionsoft}
&\frac{d}{d\,\ln\,\nu} {\cal M}_{S,+}(\ell,M,\mu_M,\nu) \nn \\
&= \bigg[\frac{\alpha_s}{4\pi} \, \Gamma_0 \,\ln\bigg( \frac{M^2}{\mu_M^2}  \bigg) \bigg] {\cal M}_{S,+}(\ell,M,\mu_M,\nu) \,,
\end{align}
As for the current and jet  mass mode matching coefficients, there is no
$\nu$-evolution for $\mu_M=M$. Solving Eq.~(\ref{eq:nuevolutionsoft}), which leads to the anticipated
exponentiation, and expanding out the terms that do not involve large
logarithms we obtain
\begin{align}
&\mu_S {\cal M}_S(\ell,M,\mu_M,\nu_-,\nu_+) \nn \\
&=\exp\bigg[\frac{\alpha_s} {4\pi}\, \Gamma_0\,\ln\bigg( \frac{M^2}{\mu_M^2}
\bigg) \ln\bigg( \frac{\nu_-}{\nu_+}  \bigg) \bigg]\left(\delta(\tilde{\ell})+\frac{\alpha_s
C_F}{4\pi}\right.\nn\\
&\times \left\{\delta(\tilde{\ell})\left[-4\,\ln\left(\frac{M^2}{\mu_M^2}\right) \left(\ln\left( \frac{\mu_S\nu_-}{\mu_M^2}\right)-\ln\left( \frac{\nu_+}{\mu_S}\right)\right)\right.\right. \nn\\
&\hspace{0.2cm}\left.\left.+\left.2\, \ln^2\left(\frac{M^2}{\mu_M^2}\right)+\frac{\pi^2}{3}\right]-\left[\frac{\theta(\tilde{\ell})}{\tilde{\ell}}\right]_{+} 8\,\ln\left(\frac{M^2}{\mu_M^2}\right)\right\}\right) \,.
\end{align}
It is convenient to adopt the choices $\nu_-=\mu_M^2/\mu_S$ and
$\nu_{+}=\mu_S$, which then gives 
\begin{align}
\mu_S {\cal M}_S(\ell,M,&\mu_S,\mu_M)=\exp\bigg[\frac{\alpha_s} {4\pi}\, \Gamma_0\,\ln\bigg( \frac{M^2}{\mu_M^2}
\bigg) \ln\bigg( \frac{\mu_M^2}{\mu_S^2}  \bigg) \bigg] \nn \\
&\times \left(\delta(\tilde{\ell})+\frac{\alpha_s
C_F}{4\pi}\left\{\delta(\tilde{\ell})\left[2\, \ln^2\left(\frac{M^2}{\mu_M^2}\right)+\frac{\pi^2}{3}\right]\right.\right.\nn\\
&\left.\left.-\left[\frac{\theta(\tilde{\ell})}{\tilde{\ell}}\right]_{+} 8\,\ln\left(\frac{M^2}{\mu_M^2}\right)\right\}\right) \,.
\end{align}
This agrees exactly with the product of ${\cal M}_H$ and ${\cal M}_J$ with the
assignment $\mu_S=\mu_J^2/\mu_H$,
\begin{align}
Q&\mathcal{M}_S(Q\tau,M,\mu_S,\mu_M)=\nn\\
&|\mathcal{M}_H(Q,M,\mu_H,\mu_M)|^2\times Q^2\mathcal{M}_J(Q^2\tau,M,\mu_J,\mu_M)\,,
\end{align}
 and confirms that the consistency relation
   Eq.~(\ref{eq:smoothnessconstr}) also applies beyond the fixed order
   approximation.

\section{Secondary Massive Fermions}
\label{sect:outlook}

As outlined in the introduction our main motivation for studying the effective
field theory setup involving the 
mass mode gauge bosons is the treatment of massive secondary quark-antiquark
pairs arising from the ${\cal O}(\alpha_s^2)$ gluon splitting 
diagrams of Fig.~\ref{fig:QCDdiag2}. For cases where the invariant mass of the quark pair
enters the 
observable the corresponding effects can be calculated from the mass mode gauge
boson results with the help of dispersion integrations. In this section we
outline the procedure. In our presentation we assume the existence of $n_f$
massless quark flavors and one quark species with mass $m$. 
The
presentation of calculational details for thrust and other quantities is beyond
the scope of this paper and will be given in subsequent publications.   

The vacuum polarization
$\Pi(q^2)$ describing a massive fermion 
bubble can be expressed through an integral over its absorptive part. The
unsubtracted (unrenormalized) version reads  
\begin{equation}
\Pi(q^2) = 
\frac{1}{\pi} \int_{4m^2}^\infty{dM^2 \frac{1}{M^2-q^2-i \epsilon}
\,\,\mathrm{Im} \left[\Pi(M^2)\right]} \, ,
\label{eq:dispersion1unsubtracted}
\end{equation}
while the  subtracted (on-shell and finite) version has the form
\begin{align}
\Pi^{\rm os}(q^2)& = 
\Pi(q^2)-\Pi(0) \nn\\ 
=&\frac{q^2}{\pi} \int_{4m^2}^\infty{\frac{dM^2}{M^2} \frac{1}{M^2-q^2-i \epsilon}
  \,\,\mathrm{Im} \left[\Pi(M^2)\right]} \, .
\label{eq:dispersion1subtracted}
\end{align}
The absorptive part in $d$ dimensions reads
\begin{align}
\mathrm{Im}&\left[\Pi(q^2)\right] = \theta(q^2-4m^2) \, g^2 T_f n_f
\frac{2^{3-2d}\pi^{(3-d)/2}\tilde{\mu}^{2\epsilon}}{\Gamma\left(\frac{d+1}{2}\right)}\nn\\
&\times(q^2)^{(d-4)/2}
\left(d-2+\frac{4m^2}{q^2}\right)\left(1-\frac{4m^2}{q^2}\right)^{(d-3)/2} \, .
\label{eq:imaginaryPiddim}
\end{align}
The correction of the massive quark-antiquark loop to the gluon propagator, as
illustrated in Fig.~\ref{fig:dispersion}, can be expressed in terms of dispersion relations. Denoting the external momentum by $q^\mu$ the relations read
\begin{align}
\Pi_{\mu\nu}^{\textnormal{eff}}(q^2)&\equiv\frac{(-i)^2
  g_{\mu\rho}\Pi^{\rho\sigma}(q^2)g_{\sigma\nu}}{(q^2+i \epsilon)^2}
\nn\\&=\frac{1}{\pi}\int_{4m^2}^\infty{\frac{dM^2}{q^2}\mathrm{Im}\left[ \Pi(M^2)\right]\frac{-i\left(g_{\mu\nu}-\frac{q_\mu q_\nu}{q^2}\right)}{q^2-M^2+i \epsilon}} \, .
\label{eq:effectivegluonpropagatorunsubtracted}
\end{align}
for the unsubtracted  and unrenormalized version  and 
\begin{align}
\Pi_{\mu\nu}^{\textnormal{eff,os}}(q^2)&\equiv\frac{(-i)^2
  g_{\mu\rho}\Pi^{\rho\sigma,{\rm os}}(q^2)g_{\sigma\nu}}{(q^2+i
  \epsilon)^2}\nn\\&
 =\frac{1}{\pi} \int_{4m^2}^\infty{\frac{dM^2}{M^2}\mathrm{Im} \left[\Pi(M^2)\right]\frac{-i\left(g_{\mu\nu}-\frac{q_\mu q_\nu}{q^2}\right)}{q^2-M^2+i \epsilon}} \, .
\label{eq:effectivegluonpropagatorsubtracted}
\end{align}
for the subtracted version involving the on-shell renormalized vacuum
polarization. 

The relations in
Eqs.~(\ref{eq:effectivegluonpropagatorunsubtracted}) and 
(\ref{eq:effectivegluonpropagatorsubtracted}) involve massive gluon 
propagators\footnote{The longitudinal contributions differ, but do not
contribute due to gauge invariance.} 
and provide the formal connection to the mass mode gauge boson considerations
discussed in the previous sections. To determine the corrections arising from the secondary massive
quark-antiquark pair one can first calculate the corresponding ${\cal
  O}(\alpha_s)$ diagrams with massive ``gluon'' propagators and then 
convolute the result according to
Eqs.~(\ref{eq:effectivegluonpropagatorunsubtracted}) and
(\ref{eq:effectivegluonpropagatorsubtracted}). 
We note that in general the final convolution needs to be carried out involving
entirely $d$~dimensional expressions for the absorptive part of the vacuum
polarization as well as for the ${\cal O}(\alpha_s)$ diagrams with the massive
``gluons''. 
The various forms
of the factorization theorems discussed in Sec.~\ref{sect:setup} concerning
the arrangement of matrix elements, matching conditions and renormalization
group factors remain.

Massive secondary quarks contribute to the renormalization evolution 
only in those terms of the factorization theorem dealing with scales above the matching scale
$\mu_m \sim m$. The dispersion integration method can deal with this
issue as well. Indeed, the way secondary quarks
contribute to the renormalization group evolution can be related to
the  use of the subtracted versus the unsubtracted dispersion relation:
when the massive secondary quark pair represents an active
flavor contributing to the renormalization group evolution together with the
massless quarks and gluons, it is convenient to employ the unsubtracted dispersion
relation of Eq.~(\ref{eq:effectivegluonpropagatorunsubtracted}). This is the
case for the calculations of the massive secondary quark 
pair effects to the hard Wilson coefficient in scenarios~II, III
and IV, to the jet function in scenarios~III and IV and to the soft function
in scenario~IV. Here the massive quarks contribute as dynamical degrees of
freedom and enter in the renormalization group evolution of the respective
components in the factorization theorem as well as of the strong coupling. In particular we
find that the associated UV-divergences are mass-independent and agree with the
ones obtained for massless secondary quarks (as we also found in an analogous way
for the one-loop 
corrections from the mass mode gauge bosons). 
The use of the unsubtracted dispersion relation also implements mass-dependent logarithms in
the UV-finite terms that are essential to reach the correct massless limit for $m\to
0$.  Thus in these cases the outcome concerning renormalization group
evolution for one massive secondary quark-antiquark pair is simply that the
number of active flavors is $n_f+1$, where $n_f$ is 
the number of massless quark flavors. In the factorization theorems discussed in
Sec.~\ref{sect:setup} this affects all renormalization group evolution factors
which included massless as well as massive gauge boson contributions (indicated
by the superscript $(2)$). All other renormalization group factors (indicated
by the superscript $(1)$) evolve with  $n_f$ active flavors, since they describe
evolution for scales where the massive quarks have been integrated out. 

When the massive secondary quarks are integrated out, one has to employ the
subtracted dispersion relation of
Eq.~(\ref{eq:effectivegluonpropagatorsubtracted}) with the  vacuum polarization
being in the on-shell scheme, thus giving the correct decoupling limit. In this
situation the massive quark flavor is not dynamical and does not contribute to
the renormalization group evolution. This is for example the
case for the calculations of the full-theory massive secondary quark 
pair effects to the hard Wilson coefficient in scenario~I, see Ref.~\cite{Hoang:1995ex}. The difference
between using the subtracted 
and the unsubtracted dispersion relation, which is related to the vacuum
polarization function at zero momentum transfer, $\Pi(0)$, correctly implements
the matrix element corrections related to the matching relation that connect
the strong coupling $\alpha_s$ schemes with $n_f$ and $n_f+1$ dynamical flavors, i.e. the
well-known decoupling relation.

\begin{figure}
 \subfigure[]{\epsfig{file=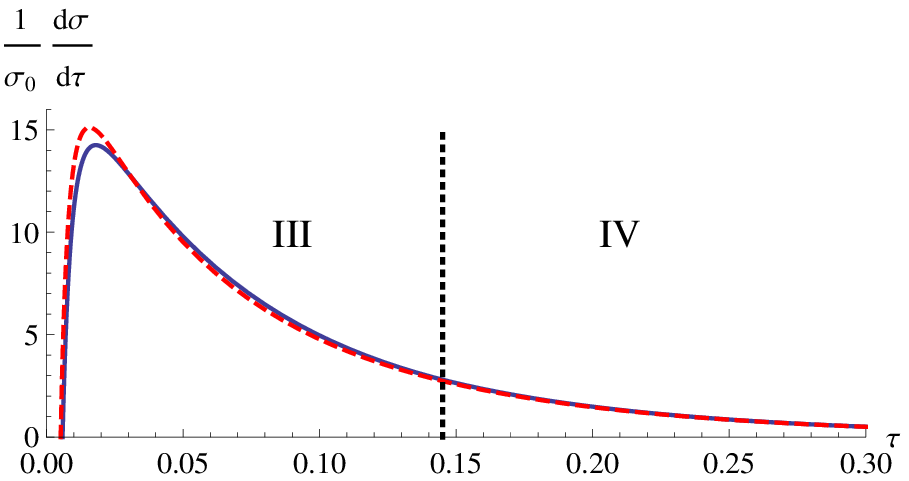,width=0.9\linewidth,clip=}}\\
 \subfigure[]{\epsfig{file=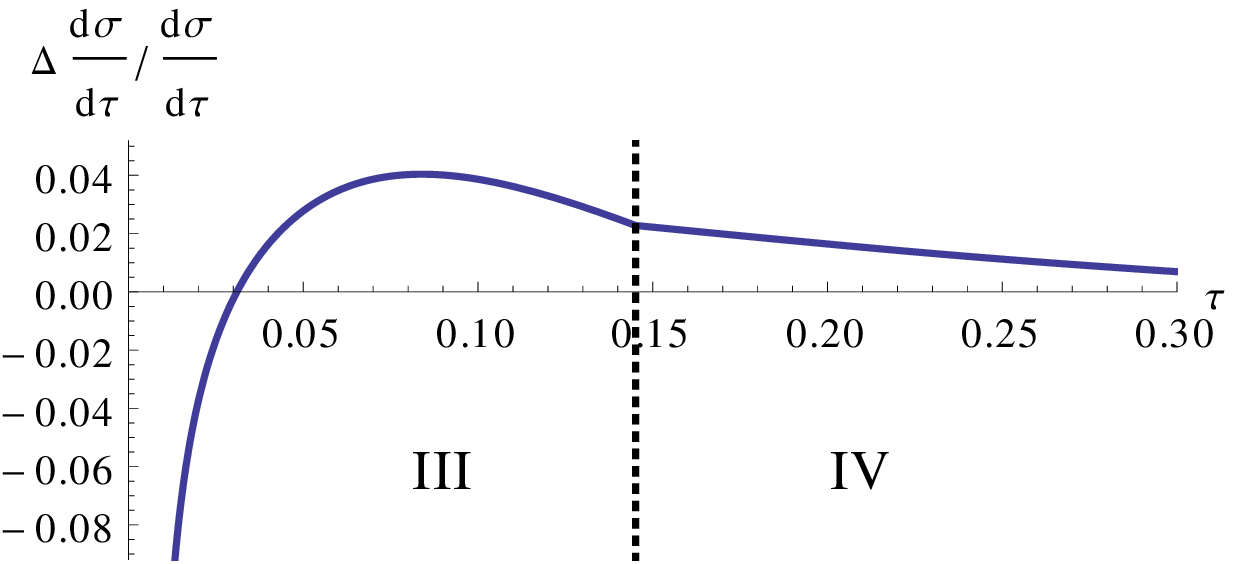,width=0.9\linewidth,clip=}}
 \caption{(a) The partonic N${}^3$LL order thrust distribution  at $Q=14$~GeV with 
   secondary bottom quark mass effects for $m_b=4.2$~GeV (blue solid
   line). The curve describes primary production of $n_f=4$ massless quarks and
   secondary production of the massless quarks and the bottom quark. 
   The red dashed line shows the result for vanishing bottom quark mass. (b)
   Relative size of secondary bottom quark mass corrections.
   The black dotted lines indicate the boundary of theory description in
   scenarios~III and IV. \label{Thrust_14}} 
\end{figure}

\begin{figure}
 \subfigure[] {\epsfig{file=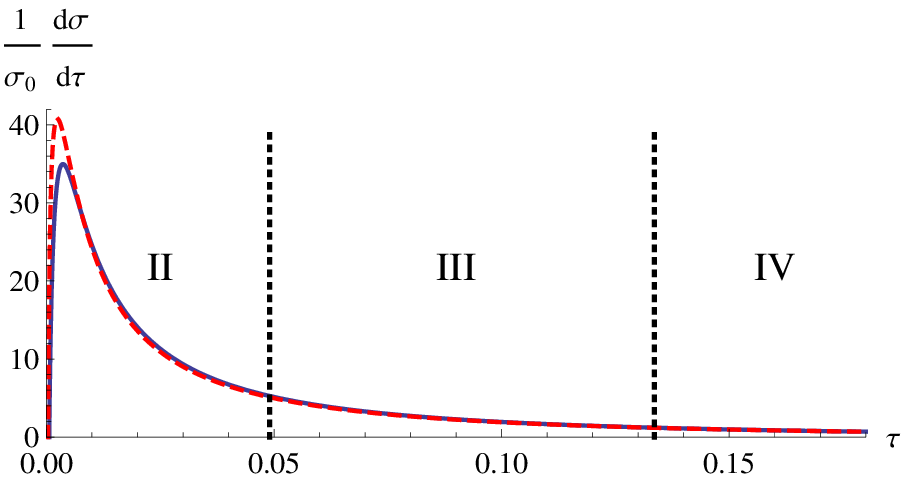,width=0.9\linewidth,clip=}}\\
 \subfigure[] {\epsfig{file=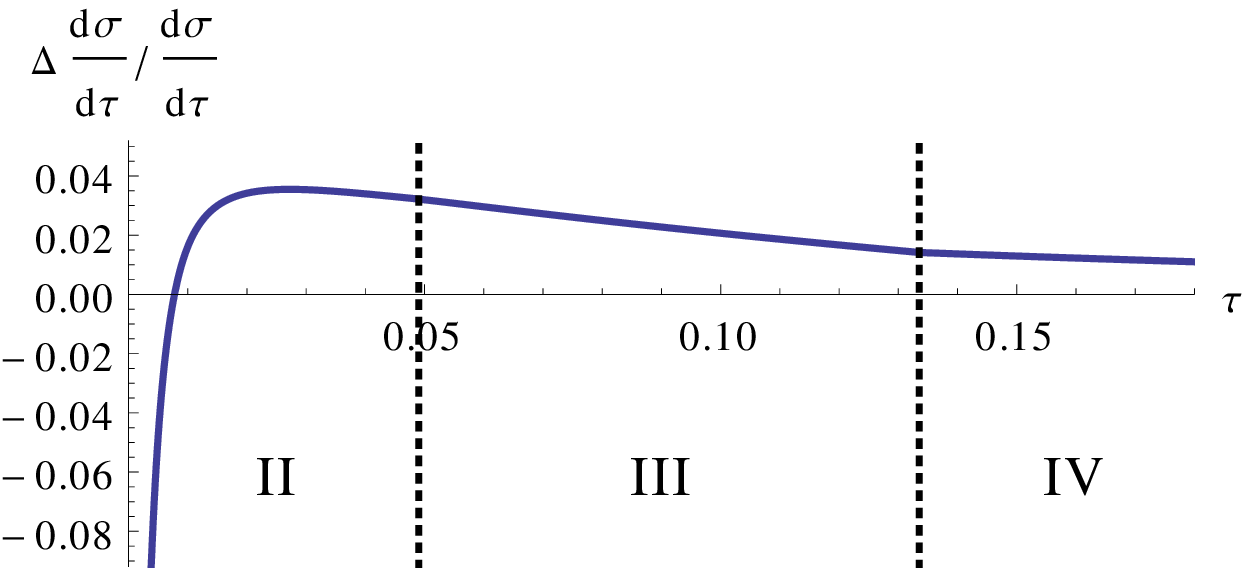,width=0.9\linewidth,clip=}}
 \caption{(a) The partonic N${}^3$LL order thrust distribution  at $Q=500$~GeV with 
   secondary top quark mass effects for $m_t=175$~GeV (blue solid
   line). The curve describes primary production of $n_f=5$ massless quarks and
   secondary production of the massless quarks and the top quark. 
   The red dashed line shows the result for vanishing top quark mass. (b)
   Relative size of secondary top quark mass corrections.
   The black dotted lines indicate the boundary of theory description in
   scenarios~II, III and IV.
\label{Thrust_500}} 
\end{figure}

We conclude with a brief discussion of the numerical
size of secondary bottom quark mass effects in the singular partonic
thrust distribution at N${}^3$LL order\footnote{We use the canonical SCET
counting, where N${}^3$LL order refers to
${\cal O}(\alpha_s^2)$ matrix elements and matching conditions and 
four-loop renormalization group evolution.}
where all two-loop mass corrections were calculated from the massive gluon
results via the dispersion method described above.\footnote{For the soft
  function the dispersion method leads to an approximate result for the two-loop
  non-logarithmic corrections which is however sufficient for the brief analysis
  carried out here.}
In Fig.~\ref{Thrust_14}a the blue solid line shows the result for the
production of $n_f=4$ massless primary quarks at $Q=14$~GeV with secondary
production of the $n_f=4$ massless quarks and the 
bottom quark with mass $m_b=4.2$~GeV. 
The red dashed line shows
the result when the bottom mass is zero. We use profile functions similar
to the ones used in Ref.~\cite{Abbate:2010xh} to describe the thrust dependence 
of the jet and soft scale. More details are provided in Ref.~\cite{Hoang:2013}. The massive effects are quite sizeable. In the peak
region the bottom mass leads to a reduction and in the tail region to an
enhancement resulting in a reduced negative slope of the distribution.
As can be seen
from Fig.~\ref{Thrust_14}b, where the relative size of the mass effects with
respect to the massless result is shown, 
the values of the thrust distributions
are enhanced up to 4\% in the dijet tail region $0.05\le\tau\le 0.15$. In the
peak region $\tau < 0.05$ the relative size of the reduction of the distribution
is even larger and leads to a sizeable effect. 
Although non-perturbative effects
have to be included for a more definite numerical analysis, it is clear that the
secondary bottom mass effects are important for the analysis of
event-shape distributions at lower c.m. energies from
JADE~\cite{MovillaFernandez:1997fr} or from current B-factory data.
In the figure we have also indicated the field theory scenarios needed to
describe the respective thrust ranges according to the hierarchies of our profile functions: for $Q=14$~GeV and $m_b=4.2$~GeV only
scenarios~III and IV can arise because for very small $\tau$ the bottom mass
is located between the soft and the jet scale, which then increase for larger $\tau$. We have implemented the factorization theorem with a strict
expansion of all two-loop corrections in the product of the matching conditions
and the jet and soft functions, and there is no displacement of the blue solid
curve at the transition from scenario~III to scenario~IV. The change of slope
visible at the transition point of scenarios~III and IV is related to the
use of the different dynamical flavor numbers for the renormalization group evolution.

An example where three scenarios are needed to describe the full thrust
spectrum is the case of secondary top quark production with $m_t=175$~GeV for $Q=500$~GeV. Here the
top mass is between the hard and the jet scale for small $\tau$, and also
scenario~II is required. The distributions with and without secondary top quark
mass effects are displayed in Fig.~\ref{Thrust_500}a using the same labeling as
in Fig.~\ref{Thrust_14}. We see that the behavior and the size of the secondary
top quark mass corrections are very similar to the bottom quark case discussed
before, albeit with the property that the peak region of the partonic
distribution is confined to much smaller $\tau$ values due to the large c.m.\
energy $Q$. A more detailed numerical analysis is presented in Ref.~\cite{Hoang:2013}.

\section{Conclusions}
\label{sect:conclusions}

In this paper we have set up a method to deal with secondary massive quark
radiation effects for inclusive jet cross sections. Depending on the mass
value
and the kinematic variables in the jet cross section, the jet invariant 
mass and the scale of soft large angle radiation can vary continuously
crossing
mass thresholds involving the secondary quark pair. This makes the
theoretical
method to deal with calculations of matrix elements, matching conditions
and the
summation of the correct large logarithmic terms complicated. We have set
up an effective
theory framework that combines the massless partonic modes described in
soft-collinear effective theory with collinear and soft mass modes. While
the collinear and soft modes for massless quarks and gluons in inclusive jets
typically have different invariant masses, the mass-shell fluctuations of collinear and soft mass
modes have
the same invariant masses. Moreover, the mass modes can also fluctuate with invariant masses like their massless counterparts if the mass is small. This leads to different effective theories
depending
on the values of the mass and the kinematic variables. We have developed
such an
effective theory scheme for thrust in $e^+e^-$ collisions considering the
effects of secondary massive quark radiation (through gluon splitting) in
massless quark pair production. 
The resulting scheme for thrust requires four different effective theories
and
allows for a continuous description from ultra-heavy quark masses in the
decoupling limit down to the limit of vanishing quark mass where the
massless description is recovered. The setup can thus be applied to every
possible kinematic situation. 

The mass mode setup is in the spirit of the variable flavor
number scheme originally developed by  Aivazis, Collins, Olness and Tung
where
infrared-safe hard Wilson coefficients were obtained with the subtraction of
low-virtuality parton splitting terms to deal with quark masses in hadron
collisions which are smaller than the hard scale but larger than
$\Lambda_{\rm
  QCD}$. Our method represents an effective field theory realization of the 
variable flavor number scheme using the framework of soft-collinear effective
theory for inclusive jet processes and its principles can be well applied to other processes
involving heavy quark production including hadron collisions. 

For thrust and other observables where the invariant mass of the massive
quark
pair enters the observable, it is possible to describe the secondary massive
quarks through a dispersion
integral involving the absorptive part of the massive quark vacuum
polarization
function and a massive ``gluon'' propagator. This allows to discuss the field
theory setup for the simpler case of a spontaneously broken non-Abelian gauge
theory combined with QCD involving massless quarks where the additional gauge
bosons have a common mass. Conceptual as well as many technical aspects
can be
studied in great detail considering these massive ``gluons''. 
In this work we have concentrated on developing the effective theory
framework and the results concerning the massive gauge bosons, and we
outlined the
calculations required to obtain the results for secondary massive quarks.
Other
applications shall be addressed in subsequent work.

\begin{acknowledgments}
 We would like to thank Guido Bell and Vicent Mateu for helpful discussions and comments on the draft. 
\end{acknowledgments}

\bibliography{references}{}
\bibliographystyle{my_bibstyle}
\end{document}